\documentclass{article}

\usepackage{hyperref}
\usepackage[T1]{fontenc}
\usepackage[utf8]{inputenc} 

\usepackage{graphicx}
\usepackage{amsmath}
\usepackage{amssymb}
\usepackage{bm,mathtools}
\usepackage{graphicx,booktabs,array}
\usepackage[section]{placeins}
\graphicspath{{figures/}}

\newcommand{\dd}{\mathrm{d}}
\newcommand{\e}{\mathrm{e}}
\newcommand{\ii}{\mathrm{i}}
\newcommand{\Tr}{\operatorname{Tr}}
\newcommand{\E}{\mathbb{E}}

\newcommand{\HH}{\mathrm H}

\begin{document}

\title{Exact Nonperturbative Equilibrium Mode Statistics in Nonlinear Wave and Lattice Systems}
\author{%
Jialin Zhang, Yong Zhang and Hong Zhao%
\thanks{Corresponding author: \href{mailto:zhaoh@xmu.edu.cn}
{\texttt{zhaoh@xmu.edu.cn}}}\\[2mm]
\small Department of Physics, Xiamen University,\\
\small Xiamen 361005, Fujian, China
}

\maketitle

\begin{abstract}
We derive exact finite-size nonperturbative representations of equilibrium modal occupations and related statistics for three broadly representative nonlinear systems: the Majda--McLaughlin--Tabak (MMT) dispersive-wave model, the Fermi--Pasta--Ulam--Tsingou (FPUT)-$\beta$ anharmonic chain, and the discrete nonlinear Schr\"odinger (DNLS) lattice field. The predicted occupations agree closely with independent numerical simulations from weak to strong nonlinearity. Nonperturbative theory can also be essential at weak coupling: in the DNLS quasicondensation regime, large low-mode occupations and coherence over large distances amplify interaction effects even when the nonlinear coefficient is very small, while the nonperturbative results remain accurate. We further show that finite-ring DNLS occupations are an exact sum of positive Rayleigh--Jeans contributions whose weights sum to the temperature. Each contribution has its own correlation length. This decomposition explains when a single Rayleigh--Jeans law applies and why it fails near quasicondensation. In MMT and DNLS, the exact occupations also determine the mean modal frequencies, even when the frequency spectra broaden or split. These nonperturbative results then allow us to test two representative perturbative approaches. An appropriate treatment of the mean interaction yields accurate low-order approximations even at strong nonlinearity. At higher orders, however, the corrections no longer decrease and successive approximations oscillate with growing amplitude. Both finite-order approximations tested near weak-coupling quasicondensation also fail markedly. Thus neither low-order agreement nor a small nonlinear coefficient ensures a reliable perturbative description. These results establish the need for nonperturbative theory in both strongly interacting and coherent-wave regimes, and provide a quantitative basis for understanding how energy, particles, and optical power are distributed among modes in nonlinear wave and lattice systems.
\end{abstract}


\section{Introduction}
\label{sec:introduction}

Nonlinear interactions allow waves to exchange energy and determine how
energy, particles, and optical power are shared among modes. When the
interaction energy becomes comparable to the dispersive or harmonic energy,
mode coupling can substantially change this distribution, shift modal
frequencies, and broaden or split frequency spectra. Strong occupation of
low-frequency modes can also make interactions important even when the
nonlinear coefficient is small. Determining the equilibrium modal
occupations in these regimes is therefore a basic problem in nonlinear
statistical physics. Most analytical approaches begin with independent
modes and treat their interactions as small corrections. When those
corrections are no longer small, convergence and finite-order accuracy
cannot be taken for granted. Nonperturbative theory is needed to determine
the interacting mode statistics without assuming weak nonlinearity.

Modal occupations connect this problem directly to experiment and
applications. In optical systems, they describe how optical power is
distributed among modes and influence the spatial profile and brightness of
a beam\cite{Krupa2017,Wu2019}. In anharmonic lattices, the mean-square modal displacements determine
the harmonic potential energy and provide basic information for studying
thermalization, fluctuations, and transport. In atomic gases, momentum distributions and condensate-number fluctuations
provide complementary probes of coherence and condensation
\cite{Jacqmin2012,KrukEtAl2025}. Mode-resolved experiments in multimode
optical fibers have observed Rayleigh--Jeans thermalization
\cite{Pourbeyram2022,Zhong2023} and optical wave condensation \cite{Sun2012,Fusaro2019,Baudin2020}.
Predicting such distributions beyond weak coupling requires a theory that
retains the correlations produced by interactions, rather than treating
each mode as statistically independent.

The weakly nonlinear limit provides a well-established starting point.
Linear normal modes, random phases, and approximately Gaussian statistics
lead to wave kinetic equations \cite{Choi2005,NewellRumpf2011} and, at equilibrium, to
Rayleigh--Jeans relations fixed by the conserved quantities
\cite{Nazarenko2011}. This description has been tested numerically in
Gross--Pitaevskii dynamics \cite{Zhu2022} and in direct and inverse cascades
of turbulent Bose gases \cite{Zhu2023,ZhuKrstulovicNazarenko2026}. The weak limit also has a rigorous
mathematical foundation. Weakly nonlinear Schr\"odinger evolution with
random initial data has been analyzed rigorously \cite{Lukkarinen2011}.
For the cubic nonlinear Schr\"odinger equation in $d\geq3$, the wave kinetic
equation has been derived in a combined large-system, vanishing-nonlinearity
limit \cite{DengHani2023}; the persistence of mode independence and the
evolution of higher-order statistics in this limit have also been
established \cite{DengHani2026}. These results provide a controlled account
of weakly nonlinear waves, but do not directly determine the occupations
of finite systems at finite or strong interaction strength.

Renormalized-wave theories account for mean interaction effects through
shifts of the modal frequencies. Such methods describe thermalized
Fermi--Pasta--Ulam--Tsingou (FPUT)-$\beta$ chains at appreciable nonlinearity
\cite{Gershgorin2005,Gershgorin2007}. In the Majda--McLaughlin--Tabak (MMT)
model, nonlinear frequency shifts modify the resonance conditions and the
Rayleigh--Jeans relation \cite{Lee2009}; interactions can even generate
effective dispersion when the linear dynamics is nondispersive
\cite{Lee2013}. Their success raises a physical question: how much of the
equilibrium mode distribution is captured by a frequency shift, and what
remains in the fluctuations around that mean? Another approach calculates
thermal averages directly from the interacting Gibbs distribution. It has
yielded important results for the thermodynamics of the discrete nonlinear
Schr\"odinger (DNLS) model \cite{Rasmussen2000,Johansson2004} and for
finite-temperature condensates on one-dimensional rings
\cite{Nunnenkamp2007}. Nevertheless, explicit predictions for individual
modal occupations at finite size and arbitrary nonlinear strength remain
scarce for widely used wave and lattice models.

Here we derive exact finite-size nonperturbative representations of
equilibrium modal occupations and related statistics for two MMT branches,
the FPUT-$\beta$ chain, and the DNLS lattice field. The MMT model captures
basic features of dispersive-wave turbulence \cite{Majda1997,Chibbaro2017};
FPUT chains are central to the study of thermalization
\cite{Onorato2015,Lvov2018,Onorato2023} and heat transport
\cite{Lepri1997,Lepri2003}; and DNLS describes nonlinear waveguide arrays
\cite{Eisenberg1998,ChenSegevChristodoulides2012} and classical fields in optical lattices
\cite{Trombettoni2001}. Together they cover dispersive waves, anharmonic
vibrations, and lattice fields with conserved particle number or optical
power. For each model, we calculate thermal averages with the full Gibbs
weight and reduce the modal occupations to integrals or transfer-operator
expressions that can be evaluated without expanding in the nonlinear
coupling. Independent Monte Carlo simulations of the original
Hamiltonians confirm the predictions from weak to strong nonlinearity.
These results determine how the modes are populated even when interaction
energy is comparable to, or larger than, the linear energy.

The DNLS results show why nonperturbative theory can also be necessary when
the nonlinear coefficient is small. Classical nonlinear waves can
accumulate strongly in low-frequency modes \cite{Connaughton2005}. In a
finite one-dimensional DNLS ring, this accumulation leads to a smooth
quasicondensation crossover as phase coherence extends across a substantial
part of the system. The nonperturbative occupations remain accurate
throughout this crossover. The large low-mode occupations amplify the
effects of interactions, so a small coupling coefficient alone does not
make the collective behavior weakly nonlinear. We also show that the
finite-ring occupations can be written exactly as a sum of positive
Rayleigh--Jeans contributions whose weights add up to the temperature.
Each contribution has its own correlation length.
This decomposition identifies when a single Rayleigh--Jeans law is
adequate and explains why the inverse occupation need not vary linearly
with the unperturbed mode frequency near quasicondensation.

The same nonperturbative occupations predict the mean modal frequencies.
This mean is the frequency averaged over the full spectrum, weighted by
the spectral intensity. Strong interactions may broaden or split a
spectrum, so its average frequency need not coincide with any individual
peak. For MMT and DNLS, an exact equilibrium identity
\cite{Tolman1918,Tolman1938} relates this average to the modal occupation,
temperature, and chemical potential. Combining this identity with the
nonperturbative occupations gives a prediction that remains accurate for
the broad and multipeaked spectra observed in the dynamics. The
equilibrium calculation thus also determines an observable property of
the time-dependent waves.

These nonperturbative results allow us to examine directly how well
perturbation theory describes the same systems. We use connected-moment
expansions \cite{Kubo1962} to compare two approaches: expansion around a
self-consistent Gaussian/Hartree state, and expansion about renormalized
modes whose frequencies include the trivial-pairing contribution.
Appropriate treatment of the mean interaction gives accurate low-order
occupations for both MMT branches, FPUT-$\beta$, and DNLS, even at strong
nonlinearity. This improvement has a physical origin: the self-consistent
frequencies already include part of the interaction effects that would
appear at higher orders in an expansion about the linear modes. The
remaining fluctuations, however, need not be small. In representative
MMT branch-I, FPUT, and DNLS calculations through fifth order, the
corrections cease to decrease and successive approximations oscillate
about the nonperturbative result with growing amplitude. Both finite-order
approximations tested near weak-coupling DNLS quasicondensation also
develop large errors. Accurate low-order results therefore do not ensure
convergence as more terms are included, and a small nonlinear coefficient
does not ensure a reliable perturbative approximation. These comparisons
show why nonperturbative results are needed both for strong interactions
and for the large low-mode populations associated with coherence.

Section~\ref{sec:nonperturbative} derives the exact finite-size occupations.
Section~\ref{sec:dnls-quasicondensation} applies the DNLS solution to
quasicondensation, and Sec.~\ref{sec:spectral-mean-frequency} relates
occupations to mean frequencies. Sections~\ref{sec:hartree}
and \ref{sec:projected-closure} examine the two perturbative approaches.
Section~\ref{sec:dnls-infrared-failure} compares them near DNLS
quasicondensation. Section~\ref{sec:conclusions-discussion} discusses
the implications. Further derivations and numerical details are given
in the Supplementary Material.

\section{Exact Nonperturbative Equilibrium Statistics}
\label{sec:nonperturbative}

\subsection{Thermal equilibrium and the nonperturbative approach}
\label{subsec:common-gibbs}

Modal second moments describe how a nonlinear equilibrium state populates
its linear modes. We use $n_k=\langle|a_k|^2\rangle$ for the complex MMT and
DNLS fields and $n_k=\langle Q_k^2\rangle$ for FPUT modal displacement
variances. The former determine norm or optical-power occupations; the latter
determine the harmonic potential energy of each mode. Interaction energies
also involve higher-order correlations. At thermal equilibrium these
quantities are averages over the Gibbs measure. With $X$ denoting the
microscopic variables,

\begin{equation}
 \begin{gathered}
 \langle\mathcal O\rangle
 =\frac1Z\int \mathcal O(X)\,
 \exp[-\beta\mathcal H_G(X)]\,\dd X,\\
 Z=\int\exp[-\beta\mathcal H_G(X)]\,\dd X,\qquad
 \mathcal H_G=H-\sum_\alpha\lambda_\alpha I_\alpha .
 \end{gathered}
 \label{eq:full-gibbs}
\end{equation}
where $\beta=T^{-1}$, $I_\alpha$ denotes any additional invariant included in
the ensemble, and $\lambda_\alpha$ is its conjugate parameter.  Thus
$\mathcal H_G=H$ for the canonical ensembles used below, whereas
$\mathcal H_G=H-\mu\mathcal N$ for the grand-canonical DNLS ensemble.

The Gibbs weight contains the interaction at its full strength.
We seek changes of variables that make the thermal averages directly
computable, without expanding this weight in the nonlinear coefficient.

The useful variables differ between the models. For MMT, we separate
the overall field amplitude from the relative amplitudes and phases of
the modes. The radial integral and, when needed, the common phase integral
can then be performed exactly. For FPUT, the bond extensions make the
potential local. The fixed endpoints leave one length constraint, whose
Fourier representation reduces the calculation to one-dimensional
integrals. For DNLS, nearest-neighbor coupling gives a transfer operator.
Fourier expansion in the phase separates this operator into angular sectors.
Keeping these sectors yields the correlation function on the complete ring.

The resulting formulas are exact representations of the occupations and
related equilibrium averages. Their remaining integrals and eigenvalue
problems are evaluated numerically. Independent Monte Carlo (MC) sampling
of the original Hamiltonians tests the predictions without fitted parameters.
Throughout, we consider thermal equilibrium at positive temperature;
no weak-nonlinearity assumption is used.

\subsection{Finite-mode cubic-quartic MMT family}
\label{subsec:mmt}

The first reduction uses the homogeneous amplitude structure of the MMT
interaction.  Both MMT cases studied below belong to one finite positive-mode family
\cite{Majda1997}:
\begin{align}
 H={}&\sum_{k=1}^{M}k^d|a_k|^2
 +\alpha\!\sum_{\substack{i,j\geq1\\i+j\leq M}}
 \left(a_i a_j a_{i+j}^*+a_{i+j}a_i^*a_j^*\right)
 \notag\\
 &+\frac b2
 \sum_{\substack{i,j,r,s=1\\i+j=r+s}}^{M}
 (ijrs)^{c/4}a_i a_j a_r^*a_s^* .
 \label{eq:mmt-hamiltonian}
\end{align}
Here $d$ specifies the linear dispersion $\omega_k=k^d$, $\alpha$ and $b>0$
are the cubic and quartic interaction strengths, and $c$ specifies the
wave-number weight of the quartic vertex.  The two parameter branches are
\begin{center}
\begin{tabular}{c c c c l}
\toprule
branch & $d$ & $c$ & cubic coefficient & interaction content\\
\midrule
I & $1$ & $2$ & $\alpha=0$ & purely quartic/four-wave\\
II & $1$ & $0$ & $\alpha^2=b$ & mixed cubic-quartic\\
\bottomrule
\end{tabular}
\end{center}
Both branches preserve the translation-induced phase symmetry
$a_k\mapsto\e^{\ii k\chi}a_k$; branch I additionally preserves the norm
$U(1)$ symmetry $a_k\mapsto\e^{\ii\phi}a_k$.  The corresponding conserved
quantities have zero conjugate multipliers in the canonical states considered
here.

The same change of variables normalizes the quadratic energy in both branches:
\begin{equation}
 q_k=k^{d/2}a_k .
 \label{eq:mmt-q}
\end{equation}
Its Jacobian is constant and cancels from normalized averages.  The Hamiltonian
becomes
\begin{equation}
 H=R(q)+2\alpha\operatorname{Re}S_d(q)+\frac b2V_{c,d}(q),
 \qquad R(q)=\sum_k|q_k|^2,
 \label{eq:mmt-homogeneous}
\end{equation}
where
\begin{equation}
 S_d(q)=\sum_{\substack{i,j\geq1\\i+j\leq M}}
 \frac{q_iq_jq_{i+j}^*}{[ij(i+j)]^{d/2}},
 \label{eq:mmt-s}
\end{equation}
and, with $w_k=k^{(c-2d)/4}$,
\begin{equation}
 B_\ell(q)=\sum_{i+j=\ell}w_iw_jq_iq_j,
 \qquad
 V_{c,d}(q)=\sum_{\ell=2}^{2M}|B_\ell(q)|^2.
 \label{eq:mmt-v}
\end{equation}
The quartic functional is positive away from the origin.  Indeed, the
$B_\ell$ are the coefficients of
$(\sum_{k=1}^{M}w_kq_kz^k)^2$.  If every $B_\ell$ vanished, this squared
polynomial and hence every $q_k$ would vanish.  Consequently $V_{c,d}(u)>0$ on the
compact unit sphere and the positive quartic term normalizes the Gibbs measure
for any finite $\alpha$.

Write
\begin{equation}
 q=\sqrt T\,ru,
 \qquad \sum_{k=1}^{M}|u_k|^2=1,
 \qquad \dd^{2M}q=T^M r^{2M-1}\dd r\,\dd\Omega(u),
 \label{eq:mmt-polar}
\end{equation}
and define
\begin{equation}
 \eta=\alpha\sqrt{T},
 \qquad \gamma=bT.
 \label{eq:mmt-couplings}
\end{equation}
Homogeneity gives
\begin{equation}
 \frac HT=r^2+2\eta r^3\operatorname{Re}S_d(u)
 +\frac{\gamma}{2}r^4V_{c,d}(u).
 \label{eq:mmt-scaled-h}
\end{equation}

\subsubsection{Branch I: quartic interaction}
\label{subsubsec:mmt1}

For branch I, $w_k=1$ and $\eta=0$.  Denote
\begin{equation}
 V_I(u)=\sum_{\ell=2}^{2M}
 \left|\sum_{i+j=\ell}u_i u_j\right|^2.
 \label{eq:mmt1-v}
\end{equation}
If $U_u(x)=\sum_{k=1}^{M}u_k\e^{\ii kx}$, Parseval's identity yields
\begin{equation}
 \frac{1}{2\pi}\int_0^{2\pi}|U_u|^2\dd x=1,
 \qquad
 V_I=\frac{1}{2\pi}\int_0^{2\pi}|U_u|^4\dd x.
 \label{eq:mmt1-parseval}
\end{equation}
Cauchy-Schwarz therefore gives $V_I\geq1$, so the remaining sphere integral
is compact and nonsingular.

Define the radial kernel
\begin{equation}
 J_\nu(V;\gamma)=\int_0^\infty r^{2\nu-1}
 \exp\!\left(-r^2-\frac{\gamma V}{2}r^4\right)\dd r.
 \label{eq:mmt1-j}
\end{equation}
It has the closed form
\begin{equation}
 J_\nu(V;\gamma)=
 \frac{\Gamma(\nu)}{2(\gamma V)^{\nu/2}}
 \exp\!\left(\frac{1}{4\gamma V}\right)
 D_{-\nu}\!\left(\frac{1}{\sqrt{\gamma V}}\right),
 \qquad \gamma V>0,
 \label{eq:mmt1-j-closed}
\end{equation}
with $J_\nu(V;0)=\Gamma(\nu)/2$.  Let $p_k=|u_k|^2$ and let $\E_u$ denote the
normalized uniform sphere average.  The denominator contains $J_M$, whereas
inserting $|a_k|^2=(T/k)r^2p_k$ changes it to $J_{M+1}$.  Therefore
\begin{equation}
 n_k^{I}=\langle|a_k|^2\rangle
 =\frac{T}{k}
 \frac{\E_u[p_kJ_{M+1}(V_I(u);\gamma)]}
 {\E_u[J_M(V_I(u);\gamma)]}.
 \label{eq:mmt1-exact}
\end{equation}
At $\gamma=0$, this reduces exactly to $n_k=T/k$.  Equation
\eqref{eq:mmt1-exact} retains the full directional correlation between $p_k$
and the quartic convolution geometry.

\subsubsection{Branch II: cubic and quartic interactions}
\label{subsubsec:mmt2}

For branch II, $d=1$, $c=0$, and $\eta^2=\gamma=bT$.  The cubic term breaks the
norm $U(1)$ symmetry, but the associated phase remains a single integration
coordinate.  Under $u\mapsto\e^{\ii\phi}u$,
\begin{equation}
 S_1(u)\mapsto\e^{\ii\phi}S_1(u),
 \qquad V_{0,1}(u)\mapsto V_{0,1}(u),
 \qquad p_k\mapsto p_k.
 \label{eq:mmt2-phase-charge}
\end{equation}
For a fixed direction apart from its common phase let $s=|S_1(u)|$.  Choosing the phase origin
so that $S_1=s$ and integrating the collective angle gives the exact identity
\begin{equation}
 \frac{1}{2\pi}\int_0^{2\pi}
 \exp[-2\eta r^3s\cos\phi]\,\dd\phi
 =I_0(2|\eta|sr^3).
 \label{eq:mmt2-phase-integral}
\end{equation}
The modified Bessel function $I_0$ retains the full statistical weight of
the collective phase selected by the cubic interaction.

Define the mixed radial kernel
\begin{equation}
 K_\nu(s,V;\eta,\gamma)=
 \int_0^\infty r^{2\nu-1}
 \exp\!\left(-r^2-\frac{\gamma V}{2}r^4\right)
 I_0(2|\eta|sr^3)\,\dd r.
 \label{eq:mmt2-kernel}
\end{equation}
The quartic decay dominates the large-$r$ growth of $I_0$, so the kernel is
finite for every direction and every finite $\eta$.  As before, inserting
$|a_k|^2=(T/k)r^2p_k$ raises the radial index by one.  The exact finite-mode
occupation is
\begin{equation}
 n_k^{II}=\frac{T}{k}
 \frac{\E_u[p_kK_{M+1}(|S_1(u)|,V_{0,1}(u);\eta,\gamma)]}
 {\E_u[K_M(|S_1(u)|,V_{0,1}(u);\eta,\gamma)]}.
 \label{eq:mmt2-exact}
\end{equation}
Equation~\eqref{eq:mmt2-exact} separates the overall occupation scale
from its distribution among modes. The angular averages retain the
wave-number dependence of the interaction, while $I_0$ accounts for the
cubic coupling between amplitude and collective phase.

The cubic interaction is included to all orders: using
$I_0(2x)=\sum_{m\geq0}x^{2m}/(m!)^2$ and interchanging the nonnegative sum with
the radial integral gives
\begin{equation}
 K_\nu(s,V;\eta,\gamma)=
 \sum_{m=0}^{\infty}
 \frac{(\eta^2s^2)^m}{(m!)^2}J_{\nu+3m}(V;\gamma).
 \label{eq:mmt2-bessel-resummation}
\end{equation}
Equation~\eqref{eq:mmt2-bessel-resummation} resums every even cubic
contribution on top of the complete quartic Gibbs weight.  It also proves that
every observable unchanged by a common phase rotation is an even function of $\alpha$.

\subsubsection{Strong-nonlinearity limits}

The exact finite-mode formulas also determine the leading scale as
$g=bT\to\infty$:
\begin{equation}
 \begin{aligned}
 n_k^{I}&=\frac1k\sqrt{\frac Tb}
 \left[C_{M,k}^{(4)}+O(g^{-1/2})\right],\\
 n_k^{II}&=\sqrt{\frac Tb}
 \left[C_{M,k}^{(4)}+O(g^{-1/2})\right].
 \end{aligned}
 \label{eq:mmt-strong}
\end{equation}
Here $C_{M,k}^{(4)}$ is the finite pure-quartic sphere average.  Its explicit
form and the cancellation of the nominal $g^{-1/4}$ correction in branch II
are derived in the Supplementary Material.  Equation~\eqref{eq:mmt-strong}
states a scale relation only; every finite-$g$ curve below is evaluated from
the complete expressions \eqref{eq:mmt1-exact} and \eqref{eq:mmt2-exact}.

\begin{figure}[!htbp]
 \centering
 \includegraphics[width=0.96\linewidth]{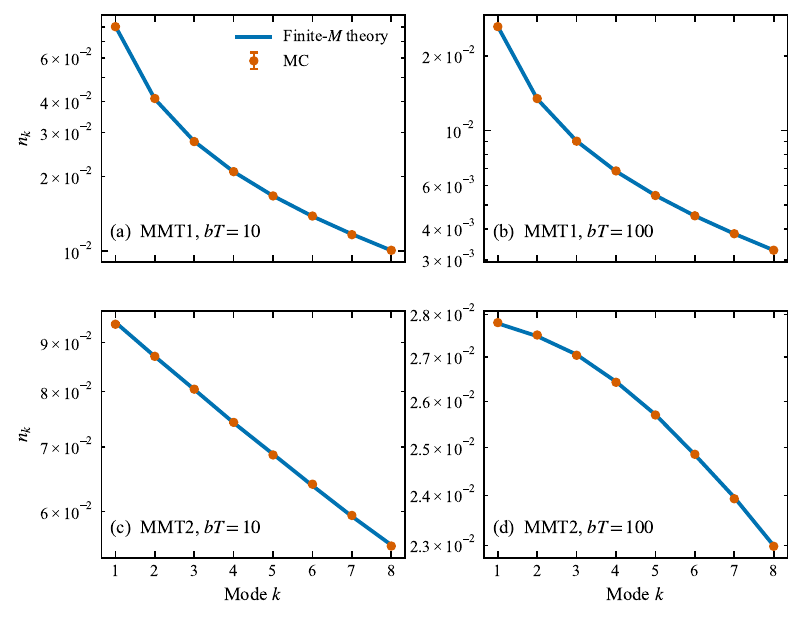}
 \caption{Nonperturbative equilibrium occupations of two branches of the
 finite-mode MMT family. Here and below, MMT1 and MMT2 denote branches I and II, respectively.  Panels (a,b) show the purely quartic branch and
 panels (c,d) the mixed cubic-quartic branch for $M=8$, $T=1$, and
 $bT=10$ or $100$.  Solid curves are the complete finite-$M$ nonperturbative
 predictions; symbols are independent Metropolis MC simulations of the
 original Hamiltonian, with replica standard errors.}
 \label{fig:mmt}
\end{figure}

Figure~\ref{fig:mmt} tests both MMT reductions for $M=8$ and $T=1$.  The
two parameter values $bT=10$ and $bT=100$ are evaluated
from Eqs.~\eqref{eq:mmt1-exact} and \eqref{eq:mmt2-exact} and compared with
independent Metropolis MC sampling of the corresponding original
Hamiltonians.  Within the numerical accuracy of the simulations,
the MC results confirm the theoretical predictions at both nonlinear
strengths.

\subsection{FPUT-\texorpdfstring{$\beta$}{beta} chain: exact constrained-bond solution}
\label{subsec:fput}

Consider the fixed-boundary FPUT-$\beta$ chain
\cite{Lvov2018,Onorato2023},
\begin{equation}
 H=\sum_{j=1}^{N}\frac{p_j^2}{2}
 +\sum_{j=0}^{N}\left[
 \frac{(q_{j+1}-q_j)^2}{2}
 +\frac b4(q_{j+1}-q_j)^4\right],
 \qquad q_0=q_{N+1}=0.
 \label{eq:fput-hamiltonian}
\end{equation}
The momentum integral is Gaussian and factorizes from the configurational
part.  The nonlinear equilibrium problem can therefore be solved entirely in
terms of the $N+1$ bonds
\begin{equation}
 r_j=q_{j+1}-q_j,
 \qquad j=0,\ldots,N.
 \label{eq:fput-bonds}
\end{equation}
The fixed endpoints impose one and only one constraint,
\begin{equation}
 \sum_{j=0}^{N}r_j=q_{N+1}-q_0=0.
 \label{eq:fput-constraint}
\end{equation}
Conversely, any bond vector satisfying Eq.~\eqref{eq:fput-constraint}
reconstructs the coordinates uniquely through
$q_j=\sum_{m=0}^{j-1}r_m$.  The coordinate-to-bond Jacobian is unity after one
bond is eliminated, or equivalently after the constraint is represented by a
delta function.

Rescale $r_j=\sqrt{T}\,x_j$ and define $g=bT$.  Apart from an overall power of
$T$, the configurational partition function becomes
\begin{equation}
 \mathcal Z_N(g)=\int_{\mathbb R^{N+1}}
 \left(\prod_{j=0}^{N}\dd x_j\right)
 \delta\!\left(\sum_{j=0}^{N}x_j\right)
 \prod_{j=0}^{N}w_g(x_j),
 \label{eq:fput-constrained-z}
\end{equation}
where
\begin{equation}
 w_g(x)=\exp\!\left(-\frac{x^2}{2}-\frac g4x^4\right).
 \label{eq:fput-weight}
\end{equation}
Thus the many-body quartic potential has become a product of identical local
anharmonic weights coupled only by the single length constraint.

Introduce the one-bond characteristic integral
\begin{equation}
 \phi_g(s)=\int_{-\infty}^{\infty}w_g(x)\e^{\ii sx}\dd x.
 \label{eq:fput-phi}
\end{equation}
Using
$\delta(X)=(2\pi)^{-1}\int_{-\infty}^{\infty}\e^{\ii sX}\dd s$ in
Eq.~\eqref{eq:fput-constrained-z} factorizes all bonds and gives the exact
one-dimensional representation
\begin{equation}
 \mathcal Z_N(g)=\frac{1}{2\pi}
 \int_{-\infty}^{\infty}\phi_g(s)^{N+1}\dd s.
 \label{eq:fput-z-1d}
\end{equation}

Because the constrained measure is invariant under permutations of all
$N+1$ bonds, its covariance has only two entries.  Define
\begin{equation}
 C_N(g)=\langle x_j^2\rangle,
 \qquad
 D_N(g)=\langle x_i x_j\rangle\quad(i\ne j).
 \label{eq:fput-cd}
\end{equation}
Multiplying the microscopic identity $\sum_jx_j=0$ by $x_i$ and averaging
yields $C_N+ND_N=0$.  Hence
\begin{equation}
 \langle x_i x_j\rangle
 =C_N\delta_{ij}-\frac{C_N}{N}(1-\delta_{ij}),
 \label{eq:fput-bond-cov-elements}
\end{equation}
or, in matrix form,
\begin{equation}
 \langle\bm x\bm x^{\mathsf T}\rangle
 =\frac{N+1}{N}C_N
 \left(I-\frac{\bm1\bm1^{\mathsf T}}{N+1}\right).
 \label{eq:fput-bond-cov}
\end{equation}

We transform this exact covariance back to the fixed-boundary normal
modes.  With
\begin{equation}
 q_j=\sqrt{\frac{2}{N+1}}
 \sum_{k=1}^{N}Q_k\sin\!\left(\frac{\pi jk}{N+1}\right),
 \label{eq:fput-normal-transform}
\end{equation}
let $D$ denote the $(N+1)\times N$ discrete-gradient matrix for which
$\bm r=D\bm Q$.  It satisfies
\begin{equation}
 D^{\mathsf T}D=\Omega^2,
 \qquad D^{\mathsf T}\bm1=0,
 \qquad
 \Omega^2=\operatorname{diag}[(\omega_1^{(0)})^2,\ldots,
 (\omega_N^{(0)})^2],
 \label{eq:fput-d-identities}
\end{equation}
with
\begin{equation}
 \omega_k^{(0)}=2\sin\!\left(\frac{k\pi}{2(N+1)}\right).
 \label{eq:fput-bare-frequency}
\end{equation}
Since $D$ has full column rank,
$\bm Q=\Omega^{-2}D^{\mathsf T}\bm r$.  Substituting
$\bm r=\sqrt{T}\bm x$ and Eq.~\eqref{eq:fput-bond-cov}, the projector
proportional to $\bm1\bm1^{\mathsf T}$ vanishes because
$D^{\mathsf T}\bm1=0$.  The remaining factors give
\begin{equation}
 \langle\bm Q\bm Q^{\mathsf T}\rangle
 =T\frac{N+1}{N}C_N\Omega^{-2}.
 \label{eq:fput-q-cov-matrix}
\end{equation}
Defining
\begin{equation}
 \theta_N(g)=\frac{N+1}{N}C_N(g),
 \label{eq:fput-theta}
\end{equation}
we arrive at the exact finite-size modal covariance
\begin{equation}
 \langle Q_kQ_\ell\rangle
 =\delta_{k\ell}\frac{T\theta_N(bT)}{(\omega_k^{(0)})^2},
 \qquad
 n_k\equiv\langle Q_k^2\rangle
 =\frac{T\theta_N(bT)}{(\omega_k^{(0)})^2}.
 \label{eq:fput-exact-spectrum}
\end{equation}
The inverse-square frequency dependence survives arbitrarily strong
anharmonicity because bond exchangeability and the fixed-length constraint
fix the covariance matrix. The complete nonlinear dependence is carried by
$\theta_N$. Thus each mode has the same mean harmonic potential energy
$T\theta_N/2$, although the higher connected correlations of the interacting
state need not vanish.

The scalar $C_N$ is obtained by differentiating the characteristic integral.
Since
\begin{equation}
 -\phi_g''(s)=\int_{-\infty}^{\infty}x^2w_g(x)\e^{\ii sx}\dd x,
 \label{eq:fput-phi-second}
\end{equation}
inserting one $x_j^2$ in Eq.~\eqref{eq:fput-constrained-z} gives
\begin{equation}
 C_N(g)=
 \frac{\displaystyle\int_{-\infty}^{\infty}
 [-\phi_g''(s)]\phi_g(s)^N\dd s}
 {\displaystyle\int_{-\infty}^{\infty}\phi_g(s)^{N+1}\dd s}.
 \label{eq:fput-cn-1d}
\end{equation}
Equations~\eqref{eq:fput-exact-spectrum} and \eqref{eq:fput-cn-1d} are the
complete finite-$N$ nonperturbative solution for the second-moment spectrum.

\subsubsection{Strong-nonlinearity and thermodynamic limits}

At fixed $N$, rescaling the bonds by $x_j=g^{-1/4}y_j$ gives the
strong-nonlinearity limit
\begin{equation}
 n_k^{(N)}=
 \frac{\sqrt{T/b}}{(\omega_k^{(0)})^2}
 \left[\Theta_N^{(4)}+O(g^{-1/2})\right],
 \qquad g=bT\to\infty .
 \label{eq:fput-strong}
\end{equation}
Here $\Theta_N^{(4)}$ is the finite constant generated by the constrained
pure-quartic bond measure.  Its integral representation and the subleading
term are given in the Supplementary Material.

Independently, taking $N\to\infty$ at fixed $g$ removes the effect of the
single length constraint on a finite set of local bonds.  The thermodynamic
modal spectrum is
\begin{equation}
 n_k^{(\infty)}=
 \frac{T\theta_\infty(g)}{(\omega_k^{(0)})^2},
 \qquad
 \theta_\infty(g)=\frac{1}{2g}
 \left[
 \frac{K_{3/4}(1/8g)}{K_{1/4}(1/8g)}-1
 \right].
 \label{eq:fput-theta-infty}
\end{equation}
The two limits are distinct: the first sends $g\to\infty$ at fixed size,
whereas the second sends $N\to\infty$ at fixed nonlinearity.  If the
thermodynamic expression is subsequently taken to strong nonlinearity, then
$g^{1/2}\theta_\infty\to2\Gamma(3/4)/\Gamma(1/4)$.  The Bessel reduction and
the finite-size $N^{-1}$ expansion are derived in the Supplementary Material.

Figure~\ref{fig:fput} uses $N=16$, $T=1$, and $bT=30$ or $70$.  The exact
finite-size values are $\theta_{16}=0.11751$ and $0.078890$.
Independent MC in the original displacement variables agrees with the
complete $N=16$ prediction mode by mode.  No thermodynamic or asymptotic
curve is used in this numerical validation.

\begin{figure}[!htbp]
 \centering
 \includegraphics[width=0.98\linewidth]{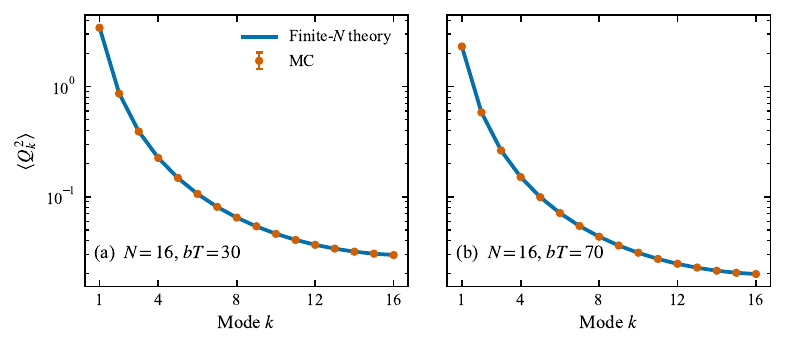}
 \caption{Finite-size nonperturbative spectrum of the fixed-boundary
 FPUT-$\beta$ chain for $N=16$, $T=1$, and $bT=30$ or $70$.  Solid curves are
 the exact finite-$N$ constrained-bond predictions, and symbols are independent
 MC data in the original displacement variables, with replica standard
 errors.}
 \label{fig:fput}
\end{figure}

\subsection{DNLS: exact correlations on a finite ring}
\label{subsec:dnls}

The periodic one-dimensional DNLS Hamiltonian and conserved norm are
\begin{equation}
 H=\sum_{j=0}^{L-1}\left[
 |\psi_{j+1}-\psi_j|^2+\frac b2|\psi_j|^4\right],
 \qquad
 \mathcal N=\sum_{j=0}^{L-1}|\psi_j|^2,
 \qquad \psi_L=\psi_0.
 \label{eq:dnls-hamiltonian}
\end{equation}
At positive temperature we use the grand-canonical measure
\begin{equation}
 \dd P=\frac{1}{Z_L}
 \exp[-\beta(H-\mu\mathcal N)]
 \prod_{j=0}^{L-1}\dd\operatorname{Re}\psi_j\,
 \dd\operatorname{Im}\psi_j .
 \label{eq:dnls-grand-canonical}
\end{equation}
Transfer-integral treatments of DNLS thermodynamics are established
\cite{Rasmussen2000,Johansson2004,Nunnenkamp2007}.  To determine the full
modal distribution at finite $L$, however, it is necessary to retain all
angular charge sectors and the way a field insertion changes those sectors.

Write
\begin{equation}
 \psi_j=\sqrt{A_j}\e^{\ii\theta_j},
 \qquad A_j\geq0.
 \label{eq:dnls-polar}
\end{equation}
Since
$\dd\operatorname{Re}\psi\,\dd\operatorname{Im}\psi
=(1/2)\dd A\dd\theta$, all constant local Jacobians cancel in normalized
averages.  Expanding the hopping term gives
\begin{equation}
 H-\mu\mathcal N
 =\sum_{j=0}^{L-1}\left[
 U(A_j)-2\sqrt{A_jA_{j+1}}
 \cos(\theta_{j+1}-\theta_j)\right],
 \label{eq:dnls-amplitude-action}
\end{equation}
where
\begin{equation}
 U(A)=(2-\mu)A+\frac b2A^2.
 \label{eq:dnls-u}
\end{equation}
The symmetric two-site transfer kernel is therefore
\begin{equation}
 \mathcal T(A,\theta;A',\theta')=
 \exp\!\left[-\frac\beta2\{U(A)+U(A')\}
 +2\beta\sqrt{AA'}\cos(\theta-\theta')\right].
 \label{eq:dnls-full-transfer}
\end{equation}

The phase enters only through a difference.  Expanding
\begin{equation}
 \e^{z\cos\phi}=\sum_{m=-\infty}^{\infty}I_m(z)\e^{\ii m\phi}
 \label{eq:dnls-bessel-expansion}
\end{equation}
separates the angular dependence into Fourier components labelled by the
integer $m$. These are the $U(1)$ sectors. The radial
operator in sector $m\in\mathbb Z$ has kernel
\begin{equation}
 K_m(A,A')=
 \exp\!\left[-\frac\beta2\{U(A)+U(A')\}\right]
 I_m(2\beta\sqrt{AA'}).
 \label{eq:dnls-km}
\end{equation}
Because $I_{-m}=I_m$, $K_{-m}=K_m$.  The exact finite-ring partition function
is
\begin{equation}
 Z_L\propto\sum_{m\in\mathbb Z}\Tr K_m^L.
 \label{eq:dnls-z-finite}
\end{equation}
The $m=0$ sector controls bulk thermodynamics as $L\to\infty$, but it is not
by itself sufficient for the momentum distribution.

Define the radial multiplication operator
\begin{equation}
 (\hat S f)(A)=\sqrt A\,f(A).
 \label{eq:dnls-s-operator}
\end{equation}
The field factors $\psi\propto\sqrt A\e^{\ii\theta}$ and
$\psi^*\propto\sqrt A\e^{-\ii\theta}$ shift the angular Fourier index by $+1$ and $-1$.  If the complementary segment of a periodic ring propagates in sector
$m$, the segment between the two insertions propagates in sector $m+1$.
Carrying out all angular integrations therefore gives the exact finite-ring
one-body correlation
\begin{equation}
 C_L(r)=\langle\psi_{j+r}\psi_j^*\rangle
 =\frac{\displaystyle\sum_{m\in\mathbb Z}
 \Tr\!\left[\hat S K_{m+1}^{r}\hat S K_m^{L-r}\right]}
 {\displaystyle\sum_{m\in\mathbb Z}\Tr K_m^L},
 \qquad 0\leq r<L.
 \label{eq:dnls-c-finite}
\end{equation}
No large-$L$ approximation has entered Eq.~\eqref{eq:dnls-c-finite}.

The content of the trace can be made explicit.  If
\begin{equation}
 K_m|m,\alpha\rangle=\lambda_{m\alpha}|m,\alpha\rangle,
 \label{eq:dnls-eigenproblem}
\end{equation}
then insertion of complete eigenbases on both sides of each $\hat S$ gives
\begin{equation}
 C_L(r)=
 \frac{\displaystyle\sum_{m,\alpha,\gamma}
 |\langle m,\alpha|\hat S|m+1,\gamma\rangle|^2
 \lambda_{m+1,\gamma}^{r}\lambda_{m,\alpha}^{L-r}}
 {\displaystyle\sum_{m,\alpha}\lambda_{m\alpha}^{L}}.
 \label{eq:dnls-c-finite-eigen}
\end{equation}
At $r=0$ this formula reduces to the finite-ring norm density
\begin{equation}
    \bar n =C_L(0)=
 \frac{\displaystyle\sum_m\Tr[A K_m^L]}
 {\displaystyle\sum_m\Tr K_m^L}.
 \label{eq:dnls-density-finite}
\end{equation}

With the discrete Fourier convention
\begin{equation}
 a_k=\frac{1}{\sqrt L}\sum_{j=0}^{L-1}\psi_j\e^{-\ii q_kj},
 \qquad q_k=\frac{2\pi k}{L},
 \label{eq:dnls-fourier}
\end{equation}
translation invariance gives
\begin{equation}
 n_k^{(L)}=\langle|a_k|^2\rangle
 =\sum_{r=0}^{L-1}\e^{-\ii q_kr}C_L(r).
 \label{eq:dnls-n-finite}
\end{equation}
Equations~\eqref{eq:dnls-km}, \eqref{eq:dnls-c-finite}, and
\eqref{eq:dnls-n-finite} form the exact finite-size nonperturbative solution.
They retain correlations along both directions around the ring and all
angular sectors of the periodic Gibbs measure.

\subsubsection{Strong-nonlinearity limits}

At fixed $T$, fixed $\mu$, and fixed $L$, the strong-nonlinearity limit is
\begin{equation}
 n^{(L)}(q)=\sqrt{\frac Tb}
 \left[\sqrt{\frac{2}{\pi}}+O((bT)^{-1/2})\right],
 \qquad bT\to\infty .
 \label{eq:dnls-strong}
\end{equation}
The leading spectrum is flat because the independent on-site quartic measure
dominates the hopping term.  This limit changes if $\mu$ is scaled with $b$ to
hold the density fixed.  The fixed-$\mu$ derivation and the first wave-number
dependent correction are given in the Supplementary Material.

\subsubsection{Thermodynamic limits}
The thermodynamic limit is instead taken as $L\to\infty$ at fixed
nonlinearity.  Let $\lambda_0$ be the largest eigenvalue of $K_0$, and let
$\lambda_{1\alpha}$ denote the eigenvalues of $K_1$.  Define
\begin{equation}
 \rho_\alpha=\frac{\lambda_{1\alpha}}{\lambda_0},
 \qquad
 W_\alpha=|\langle0|\hat S|1,\alpha\rangle|^2.
 \label{eq:dnls-rho-w}
\end{equation}
Then $0\leq\rho_\alpha<1$, $W_\alpha\geq0$, and the thermodynamic correlation
and spectrum are
\begin{equation}
 C(r)=\sum_{\alpha\geq0}W_\alpha\rho_\alpha^{|r|},
 \qquad
 \xi_\alpha=-\frac{1}{\ln\rho_\alpha}.
 \label{eq:dnls-c-thermo}
\end{equation}
\begin{equation}
 n_\infty(q)=\sum_{\alpha\geq0}W_\alpha
 \frac{1-\rho_\alpha^2}
 {1-2\rho_\alpha\cos q+\rho_\alpha^2}.
 \label{eq:dnls-poisson-sum}
\end{equation}
The thermodynamic-limit spectrum is therefore a positive sum of lattice
Poisson kernels, not generically a single Rayleigh-Jeans (RJ) distribution.

Writing the bare lattice frequency as
\begin{equation}
 \omega^{(0)}(q)=2-2\cos q,
 \label{eq:dnls-bare-frequency}
\end{equation}
one channel can be rearranged as
\begin{equation}
 W_\alpha\frac{1-\rho_\alpha^2}
 {1-2\rho_\alpha\cos q+\rho_\alpha^2}
 =\frac{T_\alpha^*}{\omega^{(0)}(q)-\mu_\alpha^*},
 \label{eq:dnls-one-rj}
\end{equation}
where
\begin{equation}
 T_\alpha^*=W_\alpha\frac{1-\rho_\alpha^2}{\rho_\alpha},
 \qquad
 \mu_\alpha^*=-\frac{(1-\rho_\alpha)^2}{\rho_\alpha}.
 \label{eq:dnls-effective-parameters}
\end{equation}

\begin{figure}[!htbp]
 \centering
 \includegraphics[width=0.96\linewidth]{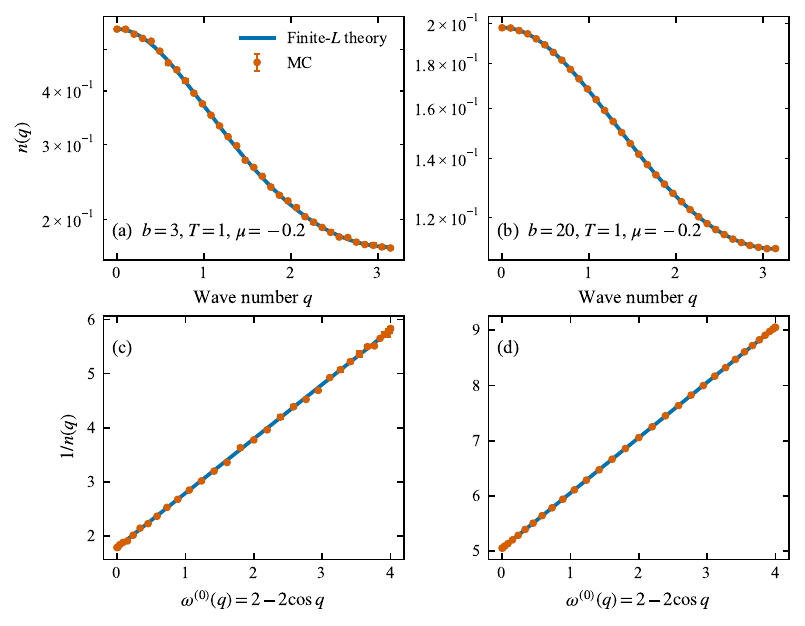}
 \caption{DNLS occupations in the normal regime for $T=1$, $\mu=-0.2$,
 $L=256$, and $b=3$ or $20$.  The upper panels show $n(q)$ against wave number.
 The lower panels show $1/n(q)$ against the bare lattice frequency.  Solid
 curves are the exact finite-ring nonperturbative predictions, and symbols are
 independent MC data with replica standard errors.}
 \label{fig:dnls-normal}
\end{figure}

Figure~\ref{fig:dnls-normal} compares the exact finite-ring occupations
with MC at $T=1$, $\mu=-0.2$, $L=256$, and $b=3$ or $20$.
The agreement holds across the wave-number range. The lower panels show
an almost linear relation between inverse occupation and bare frequency.
Section~\ref{sec:dnls-quasicondensation} explains this behavior and its
limits near quasicondensation.

\section{Nonperturbative DNLS Statistics across the Quasicondensation Crossover}
\label{sec:dnls-quasicondensation}

Small nonlinear coefficients do not always lead to nearly independent
modes. In the DNLS ring, large low-mode populations can sustain correlations
over much of the system. The finite-ring solution allows us to follow this
regime without assuming weak interactions between modes. We first test its
accuracy across the quasicondensation crossover, then use its correlation
structure to determine when a single Rayleigh--Jeans relation applies.

\subsection{Finite-ring crossover and nonperturbative accuracy}
\label{subsec:dnls-ir-finite-ring}

Using the Fourier convention and one-body correlation function introduced in
Sec.~\ref{subsec:dnls}, we monitor the crossover through
\begin{equation}
 \bar n=C_L(0)=\frac1L\sum_{k=0}^{L-1}n_k,
 \qquad
 n_0=\sum_{r=0}^{L-1}C_L(r),
 \qquad
 f_0=\frac{n_0}{L\bar n}.
 \label{eq:dnls-ir-observables}
\end{equation}
Here $\bar n$ is the mean norm density, while $n_0$ and $f_0$ resolve the
accumulation of norm in the infrared sector.  A complementary coherence
diagnostic follows from the largest transfer eigenvalues in the $m=0$ and $m=1$ sectors,
\begin{equation}
 \rho_0=\frac{\lambda_{1,0}}{\lambda_{0,0}},
 \qquad
 \xi_0=-\frac{1}{\ln\rho_0},
 \qquad
 X=\frac{\xi_0}{L}.
 \label{eq:dnls-ir-correlation-length}
\end{equation}
When $X\ll1$, correlations decay before traversing the periodic system.  As
$X$ grows, the wrap-around factors $\rho_0^L=\exp(-L/\xi_0)$ retained by the
finite-ring formula affect the lowest modes and $f_0$ rises rapidly.  At
positive temperature and finite $L$, this is a smooth quasicondensation
crossover, not a sharp finite-temperature transition in a one-dimensional
short-range system.  We therefore identify it through the joint growth of
$\xi_0/L$, $n_0$, and the low-wave-number peak rather than by assigning a
critical value to one observable.  Momentum distributions provide an
established probe of analogous one-dimensional quasicondensation crossovers.

The finite-ring solution remains accurate as this low-frequency peak grows.
At $T=1$, $L=256$, and $b=0.1$, the largest relative difference between
MC and the nonperturbative density is below $9.3\times10^{-4}$ across seven
sampled chemical potentials. At $\mu=2$, the density and zero-mode
occupation are
\begin{equation}
 \begin{aligned}
 \bar n^{\mathrm{NP}}&=19.8190,
 &\bar n^{\mathrm{MC}}&=19.8184\pm0.0006,\\
 n_0^{\mathrm{NP}}&=2769.5,
 &n_0^{\mathrm{MC}}&=2769.3\pm8.3,
 \end{aligned}
 \label{eq:dnls-ir-np-mc-numbers}
\end{equation}
The agreement therefore includes both the total norm and its large
concentration in the zero mode. The transfer eigenvalues give
\begin{equation}
 \rho_0=0.987114,
 \qquad
 \xi_0=77.10,
 \qquad
 \frac{\xi_0}{L}=0.3012,
 \qquad
 \rho_0^L\simeq0.036,
 \label{eq:dnls-quasi-scales}
\end{equation}
so correlations extend over about a third of the ring. Contributions that
propagate around the periodic boundary are no longer negligible.
The full finite-ring expression retains these contributions and agrees
with MC even for the strongly enhanced zero mode. The complete occupation
curves are compared with perturbative approximations in
Sec.~\ref{sec:dnls-infrared-failure}.

\subsection{Nonlinearity dependence of the crossover}
\label{subsec:dnls-ir-nonlinearity-dependence}

\begin{figure}[!htbp]
 \centering
 \includegraphics[width=0.98\linewidth]{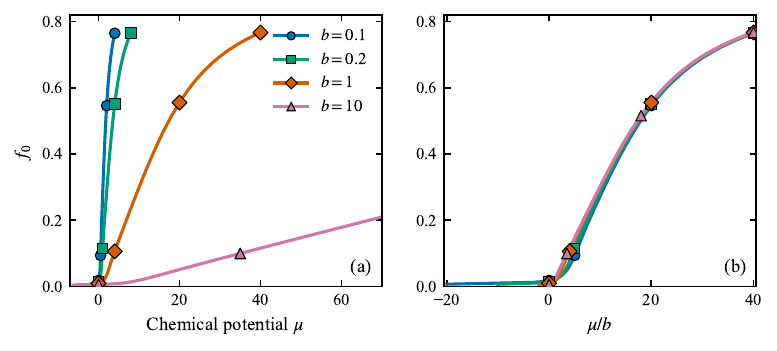}
 \caption{Finite-size DNLS quasicondensation at $T=1$ and $L=256$.
 (a) Zero-mode fraction $f_0=n_0/\sum_k n_k$ versus chemical potential for
 $b=0.1$, $0.2$, $1$, and $10$.  (b) The same data plotted against $\mu/b$.
 Solid curves are the exact finite-ring transfer-operator predictions and
 symbols are independent MC results.}
 \label{fig:dnls-ir-f0}
\end{figure}

Figure~\ref{fig:dnls-ir-f0} compares $b=0.1$, $0.2$, $1$, and $10$.
The two weak-coupling values resolve the rapid change of the low-mode
population in this regime. At fixed chemical potential, increasing $b$
reduces the zero-mode fraction over the crossover interval; the rise of
$f_0$ shifts to larger $\mu$ and becomes more gradual. The exact
finite-ring prediction follows MC throughout the scan.

The same data align more closely when plotted against $\mu/b$. This ratio
has a simple amplitude interpretation: the uniform minimum of
$-\mu|\psi|^2+(b/2)|\psi|^4$ has $|\psi|^2=\mu/b$ for $\mu>0$.
It is not the exact finite-temperature density, but identifies the scale
on which the amplitude grows along this path. Writing
$\psi_j=\sqrt{A_j}\,\e^{\ii\theta_j}$ also shows that the phase coupling is
$2\sqrt{A_jA_{j+1}}$. Thus changing $b$ at fixed $\mu$ changes both density
and phase stiffness. The broadening of the crossover in Fig.~\ref{fig:dnls-ir-f0}
describes this path, rather than a comparison at fixed density.

\subsection{Finite-ring multichannel Rayleigh--Jeans structure}
\label{subsec:dnls-ir-rayleigh-jeans-structure}

Pairing each transition $(m,\alpha)\to(m+1,\gamma)$ with its
partner with the opposite angular index and performing the finite Fourier sum reduces
Eq.~\eqref{eq:dnls-n-finite} exactly to
\begin{equation}
 n_k^{(L)}=\sum_c
 \frac{\Theta_c^{(L)}}{\omega^{(0)}(q_k)+\Delta_c},
 \qquad \omega^{(0)}(q_k)=2-2\cos q_k,
 \label{eq:dnls-finite-multirj}
\end{equation}
where $c=(m,\alpha,\gamma)$ labels a paired transition with $m\geq0$.
Write $\lambda_{c,>}$ and $\lambda_{c,<}$ for the larger and smaller
eigenvalues of the adjacent sectors, and set
$\rho_c=\lambda_{c,<}/\lambda_{c,>}$,
$\mathcal Z_L=\sum_{m,\alpha}\lambda_{m\alpha}^L$, and
$s_c=\langle m,\alpha|\hat S|m+1,\gamma\rangle$. The explicit coefficients are
\begin{equation}
 \Delta_c=\frac{(1-\rho_c)^2}{\rho_c},\qquad
 \Theta_c^{(L)}
 =\frac{|s_c|^2\lambda_{c,>}^{L}}{\mathcal Z_L}
 \frac{(1-\rho_c^L)(1-\rho_c^2)}{\rho_c}.
 \label{eq:dnls-finite-channel-weight}
\end{equation}
Each eigenvalue ratio defines a correlation length
$\xi_c=-1/\ln\rho_c$; for a channel with a long correlation length,
$\Delta_c\simeq\xi_c^{-2}$. Degenerate eigenvalues are treated by the
continuous limit of their contribution to Eq.~\eqref{eq:dnls-finite-multirj}.

The channel weights are nonnegative and obey the exact
finite-ring sum rule
\begin{equation}
 \sum_c\Theta_c^{(L)}=T,
 \qquad
 P_c^{(L)}\equiv\frac{\Theta_c^{(L)}}{T}\geq0,
 \qquad
 \sum_cP_c^{(L)}=1.
 \label{eq:dnls-finite-theta-sum}
\end{equation}
The coefficients $\Theta_c^{(L)}=TP_c^{(L)}$ are the numerators of the RJ
contributions. Their sum is $T$; they are not particle-number fractions. A channel with a small
$\Delta_c$ can contribute strongly to the lowest modes even if its numerator
weight is modest.

The frequency offsets also satisfy
\begin{equation}
 \begin{gathered}
 \sum_c\mathcal N_c\Delta_c
 =2\langle H_{\rm nl}\rangle-\mu\langle\mathcal N\rangle,
 \end{gathered}
 \label{eq:dnls-finite-mass-moment}
\end{equation}
where $\mathcal N_c\equiv\sum_k
\frac{\Theta_c^{(L)}}{\omega^{(0)}(q_k)+\Delta_c}$,
$H_{\rm nl}=\frac b2\sum_j|\psi_j|^4$. The channel definitions and both sum rules are derived in the Supplementary
Material.

The resulting occupation can thus be resolved into contributions with
different correlation lengths:

\begin{equation}
 n_k=\sum_c n_{k,c},
 \qquad
 n_{k,c}=\frac{\Theta_c^{(L)}}{\omega_k^{(0)}+\Delta_c},
 \qquad
 \sum_c\Theta_c^{(L)}=T.
 \label{eq:dnls-ir-channel-definition}
\end{equation}
If one channel supplies both the total weight and the total occupation,
Eq.~\eqref{eq:dnls-finite-theta-sum} gives $\Theta_c\to T$.
The spectrum then becomes a single RJ law
\begin{equation}
 n(q)\longrightarrow
 \frac{T}{\omega^{(0)}(q)+\Sigma_E-\mu},
 \qquad
 \Sigma_E=\frac{2\langle H_{\rm nl}\rangle}{\langle\mathcal N\rangle}
 =\frac{2h_{\rm nl}}{\bar n},
 \label{eq:dnls-principal-rj}
\end{equation}
where $h_{\rm nl}=\langle H_{\rm nl}\rangle/L$ and
$\bar n=\langle\mathcal N\rangle/L$.  Hence the physical nonlinear
frequency shift is twice the nonlinear-energy density divided by the norm
density, while the corresponding RJ offset is
$\Delta_{c}=\Sigma_E-\mu$.

Each channel is a lattice propagator with its own correlation
scale. To determine when their sum is a single RJ law, consider its continuous
frequency extension and define
\begin{equation}
 x=\omega^{(0)},\qquad
 \mathcal M_j(x)=\sum_c\frac{\Theta_c^{(L)}}{(x+\Delta_c)^j},
 \qquad n(x)=\mathcal M_1(x).
 \label{eq:dnls-rj-moments}
\end{equation}
Since $n'=-\mathcal M_2$ and $n''=2\mathcal M_3$,
\begin{equation}
 \frac{\mathrm d^2}{\mathrm dx^2}\frac1{n(x)}
 =\frac{2(\mathcal M_2^2-\mathcal M_1\mathcal M_3)}
        {\mathcal M_1^3}\leq0.
 \label{eq:dnls-inverse-concavity}
\end{equation}
The inequality follows from Cauchy--Schwarz and the positivity of the
weights. Equality throughout an interval requires all active $\Delta_c$ to
coincide; a single channel is a special case. One dominant channel or a narrow
range of frequency offsets can give an approximately linear inverse spectrum,
whereas distinct offsets generate downward curvature. This criterion holds
for the finite-ring weights and does not require discarding wrap-around
contributions.

For the normal-regime parameters $T=1$, $\mu=-0.2$, and $L=256$, the two cases
$b=3$ and $20$ have short correlation lengths. Their leading ratios are
$\rho_0=0.2859$ and $0.1449$, corresponding to
$\xi_0/L=0.00312$ and $0.00202$. The fractions of the thermodynamic
correlation overlap $\sum_\alpha W_\alpha$ outside its leading radial channel
are $1.32\times10^{-4}$ and $1.15\times10^{-5}$, respectively.
Consequently, the dominant-channel curve and the full result are nearly
indistinguishable, which explains why the exact finite-ring curves in the lower
panels of Fig.~\ref{fig:dnls-normal} are almost linear.

The infrared importance of a channel depends on both $\Theta_c^{(L)}$ and
$\Delta_c$: its zero-mode contribution is $\Theta_c^{(L)}/\Delta_c$.
A long correlation length can therefore compensate for a small numerator
weight. This distinction is essential when interpreting channel populations.

At $T=1$, $L=256$, $b=0.1$, and $\mu=2$, the $m=0$ family carries
$93.26\%$ of the total numerator weight and $98.76\%$ of $n_0$.  The soft
$(0,0,0)$ channel alone contributes $98.76\%$ of $n_0$, while the hard radial
channel $(0,0,1)$ has a comparable numerator weight but mainly affects finite
frequency.  At $\mu=10$, the $m=0$ and $m>0$ families carry $45.22\%$ and
$54.78\%$ of the numerator weight and contribute $74.18\%$ and $25.82\%$ of
$n_0$, respectively.  Several angular sectors are then required, although the long-range
contributions from their lowest radial states still dominate the zero mode.

\begin{figure}[!htbp]
 \centering
 \includegraphics[width=0.98\linewidth]{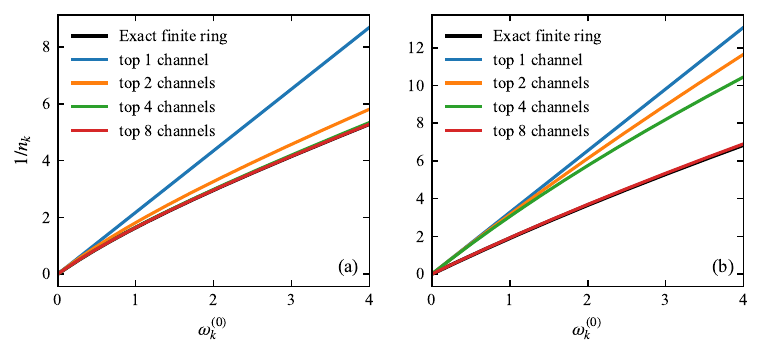}
 \caption{Cumulative-channel reconstruction of the exact finite-ring DNLS
 inverse spectrum at $T=1$, $L=256$, and $b=0.1$, plotted against
 $\omega_k^{(0)}=2-2\cos q_k$.  Channels are ranked by decreasing
 $\Theta_c^{(L)}$.  The black curve is the exact sum over all retained
 channels; colored curves are the inverses of the cumulative occupations from
 the leading 1, 2, 4, and 8 channels.  (a) $\mu=2$.  (b) $\mu=10$.
 Successive cumulative sums reconstruct the complete occupation. Their
 different slopes mainly show the spectral weight added by successive channels;
 the weaker curvature of the full inverse spectrum is quantified by
 Eq.~\eqref{eq:dnls-inverse-concavity}.}
 \label{fig:dnls-ir-channel-lines}
\end{figure}

Figure~\ref{fig:dnls-ir-channel-lines} orders the channels by decreasing
$\Theta_c^{(L)}$ and plots the inverse of their cumulative occupation.  At
$\mu=2$, the first four channels already reproduce the exact curve closely.
At $\mu=10$, channels ranked fifth and sixth in numerator weight include the
soft $(0,0,0)$ and $(2,0,0)$ contributions; the pronounced improvement
between the four- and eight-channel sums therefore illustrates why ranking
by numerator weight alone does not determine infrared importance.  The
cumulative construction makes the approach to the exact multichannel result
visible even though the final departure from a straight line is modest.

\section{Spectral Broadening and the Robustness of the Mean Frequency}
\label{sec:spectral-mean-frequency}

The occupations describe how particles or wave intensity are distributed
among modes at equilibrium.  Strong interactions also modify the temporal response:
a narrow quasiparticle peak may broaden, become asymmetric, or split into several
local maxima.  In that situation no single peak position provides an unambiguous
mode frequency.  Nevertheless, the mean over the complete spectrum still obeys an exact relation.  We show below that the nonperturbative equilibrium occupation fixes
this mean frequency without any assumption about the spectral line shape.

\subsection{Spectral moments and an exact mean-frequency identity}
\label{subsec:sf-moment-identity}

For a complex canonical mode $a_k(t)$ in a stationary Gibbs state, define
\begin{equation}
 C_k(t)=\langle a_k(t)a_k^*(0)\rangle,
 \qquad
 S_k(\omega)=\int_{-\infty}^{\infty}C_k(t)\e^{\ii\omega t}\,\dd t .
 \label{eq:sf-correlation-spectrum}
\end{equation}
With this convention, $a_k(t)\propto\e^{-\ii\Omega_k t}$ produces a peak at
$\omega=\Omega_k$.  All frequency integrals below extend over the full real
axis.  The inverse transform at $t=0$ gives the zeroth moment,
\begin{equation}
 M_k^{(0)}\equiv\frac{1}{2\pi}\int_{-\infty}^{\infty}
 S_k(\omega)\,\dd\omega=C_k(0)=n_k,
 \qquad n_k=\langle|a_k|^2\rangle .
 \label{eq:sf-zeroth-moment}
\end{equation}
Differentiating the inverse transform before setting $t=0$ gives
\begin{equation}
 M_k^{(1)}\equiv\frac{1}{2\pi}\int_{-\infty}^{\infty}
 \omega S_k(\omega)\,\dd\omega
 =\ii\dot C_k(0)=\ii\langle\dot a_k a_k^*\rangle .
 \label{eq:sf-first-moment}
\end{equation}
Thus a dynamical spectral moment is reduced to an equal-time correlation, as in
general spectral-moment sum rules \cite{Freericks2013}.  Hamilton's equation,
$\ii\dot a_k=\partial H/\partial a_k^*$, then yields
\begin{equation}
 M_k^{(1)}=
 \left\langle a_k^*\frac{\partial H}{\partial a_k^*}\right\rangle .
 \label{eq:sf-moment-hamiltonian}
\end{equation}

For the normalizable Gibbs weight
$\exp[-\beta(H-\mu\mathcal N)]$, with $\beta=T^{-1}$ and
$\mathcal N=\sum_j|a_j|^2$, integration by parts in the complex mode variable
gives \cite{Tolman1918,Tolman1938}
\begin{equation}
 \left\langle a_k^*
 \frac{\partial(H-\mu\mathcal N)}{\partial a_k^*}\right\rangle=T .
 \label{eq:sf-gibbs-identity}
\end{equation}
Here $k_{\mathrm B}=1$ and the Gibbs boundary term vanishes.  Since
$\partial\mathcal N/\partial a_k^*=a_k$, Eqs.~\eqref{eq:sf-moment-hamiltonian}
and \eqref{eq:sf-gibbs-identity} imply
$M_k^{(1)}=T+\mu n_k$.  Combining this result with
Eq.~\eqref{eq:sf-zeroth-moment} gives the exact mean frequency
\begin{equation}
 \bar\omega_k\equiv
 \frac{\displaystyle\int_{-\infty}^{\infty}
 \omega S_k(\omega)\,\dd\omega}
 {\displaystyle\int_{-\infty}^{\infty}S_k(\omega)\,\dd\omega}
 =\mu+\frac{T}{n_k} .
 \label{eq:sf-exact-centroid}
\end{equation}
The identity holds for a stationary Gibbs state with a finite first
spectral moment and the vanishing boundary term used above. It is independent
of whether the spectrum has a narrow peak.

For the canonical MMT ensembles considered here $\mu=0$, whereas DNLS retains
the chemical potential conjugate to the conserved norm.  Inserting the exact
finite-size occupations derived in Sec.~\ref{sec:nonperturbative} therefore gives
\begin{equation}
 \bar\omega_k^{\mathrm{NP}}=
 \begin{cases}
 T/n_k^{\mathrm{NP}}, & \text{MMT},\\[2pt]
 \mu+T/n_k^{\mathrm{NP}}, & \text{DNLS}.
 \end{cases}
 \label{eq:sf-nonperturbative-centroid}
\end{equation}
The equilibrium occupation thus fixes the mean frequency without a
line-shape fit. Here the identity applies to the complex canonical modes;
the FPUT displacement variance is a different observable.

\subsection{Broadened spectra and the nonperturbative mean frequency}
\label{subsec:sf-broadened-spectra}

\begin{figure}[!htbp]
 \centering
 \includegraphics[width=\linewidth]{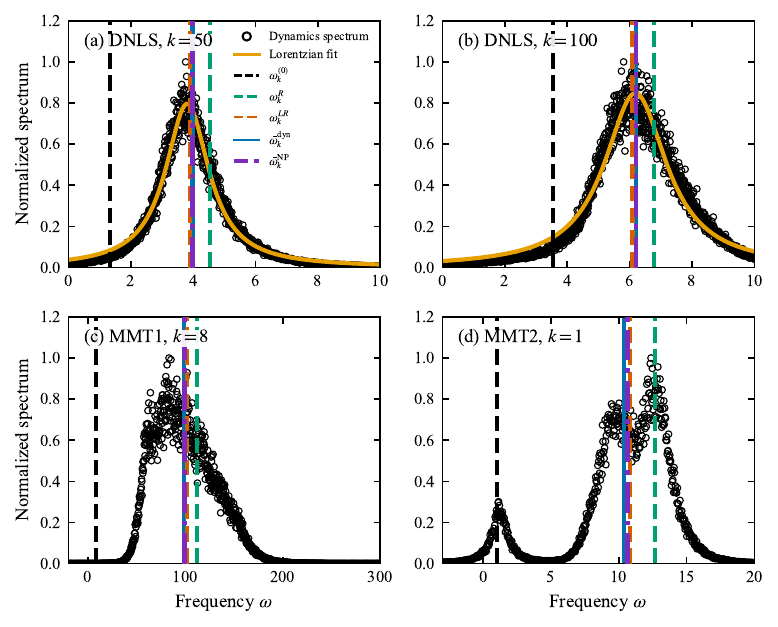}
 \caption{Spectral broadening and the nonperturbative mean frequency.
 Panels (a,b) show DNLS modes $k=50$ and $100$ for $L=256$, $T=1$, $b=7$,
 and $\mu=-0.2$.  Panel (c) shows MMT branch I for
 $(c,\alpha)=(2,0)$, $M=8$, $T=1$, $bT=10$, and $k=8$.
 Panel (d) shows MMT branch II for $(c,\alpha)=(0,\sqrt b)$,
 $M=8$, $T=1$, $bT=10$, and $k=1$.  Each spectrum is normalized by its
 maximum. Parameters are expressed in unit-temperature normalization.
 Open circles are dynamical spectra, and the smooth yellow curves
 in (a,b) are Lorentzian fits.  Vertical lines denote the harmonic frequency
 $\omega_k^{(0)}$ (black dashed), the first trivial-pairing-renormalized
 frequency $\omega_k^{\mathrm R}$ (green dashed), the leading
 residual-interaction estimate $\omega_k^{\mathrm{LR}}$ (orange dashed), the
 dynamical mean frequency $\bar\omega_k^{\mathrm{dyn}}$ (blue
 solid), and the nonperturbative prediction $\bar\omega_k^{\mathrm{NP}}$
 (purple dash-dotted line).}
 \label{fig:sf-broadening}
\end{figure}

Figure~\ref{fig:sf-broadening} compares Eq.~\eqref{eq:sf-nonperturbative-centroid}
with dynamical spectra. Three additional frequency scales are shown:
the harmonic value $\omega_k^{(0)}$, the renormalized value
$\omega_k^{\mathrm R}$, which includes the mean diagonal four-wave
interaction, and $\omega_k^{\mathrm{LR}}$, which also includes the
leading correction from the remaining interaction.
Here the mean shift is evaluated from the physical occupations, not
from a Gaussian reference. The separation of diagonal and remaining
interactions is developed in Sec.~\ref{sec:projected-closure}.
These markers show how the familiar frequency estimates sit within
the broadened spectrum. The nonperturbative mean frequency is obtained
independently from Eq.~\eqref{eq:sf-nonperturbative-centroid}.

The DNLS spectra in Figs.~\ref{fig:sf-broadening}(a) and
\ref{fig:sf-broadening}(b) are substantially broadened but remain approximately
single-peaked.  Their Lorentzian fits provide useful visual summaries of the
central regions, although no fit enters Eq.~\eqref{eq:sf-exact-centroid}.  The dynamical and nonperturbative mean-frequency markers nearly overlap on the
scale of the displayed spectra.

The MMT spectra make the distinction between a peak and a mean more explicit.
The branch-I spectrum in Fig.~\ref{fig:sf-broadening}(c) is broad and strongly
asymmetric, while the branch-II spectrum in Fig.~\ref{fig:sf-broadening}(d)
contains a low-frequency component and a broad high-frequency structure with
several local maxima.  In the latter example, $\omega_k^{(0)}$ and
$\omega_k^{\mathrm R}$ lie near two different local maxima.  These local features coexist with a different mean frequency.
In both MMT cases the nonperturbative mean frequency remains close to the
indicated dynamical mean of the complete spectrum.

The first spectral moment is therefore robust to line-shape changes that
make a single-peak description inadequate. Its exact connection to $n_k$
transfers the nonperturbative equilibrium prediction to a measurable
dynamical frequency.

\section{Self-Consistent Gaussian/Hartree Expansion}
\label{sec:hartree}

The nonperturbative formulas determine the occupations directly.
Perturbation theory serves a different purpose: it gives simpler
approximations and separates the contributions of different fluctuations.
The exact results now let us test whether adding these contributions
actually improves the occupations.

We first absorb the average interaction into a self-consistent Gaussian,
or Hartree, reference state. The remaining interaction describes
fluctuations about that state. We retain it in full, including fluctuations
of the modal intensities. The first correction vanishes when this
separation is made consistently. We then calculate the second-order
correction and examine representative sequences through fifth order.
The DNLS tests here are in the normal regime; the quasicondensation
comparison follows in Sec.~\ref{sec:dnls-infrared-failure}.

\subsection{Finite-size Hartree reference}
\label{subsec:hartree-reference}

Let $\mathcal H_G$ denote the Gibbs generator introduced in
Eq.~\eqref{eq:full-gibbs}.  Thus $\mathcal H_G=H$ for the canonical MMT and
FPUT systems and $\mathcal H_G=H-\mu\mathcal N$ for DNLS.  We write
\begin{equation}
 \mathcal H_G=\mathcal H_{\HH}+\Delta\mathcal H_{\HH}
 +\text{const.},
 \label{eq:h-decomposition}
\end{equation}
where $\mathcal H_{\HH}$ is quadratic and the residual
$\Delta\mathcal H_{\HH}$ has no quadratic Wick component.  Equivalently, the
Gaussian variational functional
$F_{\HH}+\langle\mathcal H_G-\mathcal H_{\HH}\rangle_{\HH}$ is stationary
with respect to every covariance allowed by the symmetries.  A scalar
independent of the dynamical variables cancels from normalized averages and
is suppressed in Eq.~\eqref{eq:h-decomposition}; a shift multiplying a
conserved quadratic invariant changes the covariance and must be retained.

\subsubsection{MMT}

For branch I of Eq.~\eqref{eq:mmt-hamiltonian}, use the variables
$q_k=\sqrt{k}\,a_k$ introduced in Eq.~\eqref{eq:mmt-q} and define
\begin{equation}
 \mathcal Q_4(q)=\sum_{i+j=r+s}q_iq_jq_r^*q_s^*,
 \qquad R=\sum_{k=1}^{M}|q_k|^2.
 \label{eq:h-mmt-q4}
\end{equation}
If $S_q=\sum_k\langle|q_k|^2\rangle_{\HH}$, Gaussian normal ordering gives
$\mathcal Q_4=: \mathcal Q_4:_{\HH}+4S_qR-2S_q^2$.  The Hartree reference is
therefore
\begin{align}
 \mathcal H_{\HH}^{(I)}&=A R,
 &S_q&=\frac{MT}{A},
 &A&=1+2bS_q,
 \notag\\
 A&=\frac{1+\sqrt{1+8MbT}}{2},
 &n_k^{\HH}&=\frac{T}{Ak},
 \label{eq:h-mmt1-reference}
\end{align}
and the residual is
\begin{equation}
 \Delta\mathcal H_{\HH}^{(I)}=\frac b2:
 \mathcal Q_4(q):_{\HH}.
 \label{eq:h-mmt1-residual}
\end{equation}

In branch II the quartic convolution is unweighted in the original $a_k$
variables, and contracting two fields in the cubic term gives zero because all retained
mode indices are positive.  With $S_a=\sum_k n_k^{\HH}$,
\begin{align}
 \mathcal H_{\HH}^{(II)}&=\sum_{k=1}^{M}(k+\Delta_{\HH})|a_k|^2,
 &n_k^{\HH}&=\frac{T}{k+\Delta_{\HH}},
 \notag\\
 \Delta_{\HH}&=2bS_a,
 &S_a&=T\sum_{k=1}^{M}\frac{1}{k+\Delta_{\HH}}.
 \label{eq:h-mmt2-reference}
\end{align}
The common shift $\Delta_{\HH}$ is part of the quadratic reference.  If
$H_3$ denotes the cubic term in Eq.~\eqref{eq:mmt-hamiltonian}, then
\begin{equation}
 \Delta\mathcal H_{\HH}^{(II)}=H_3+\frac b2:
 \mathcal Q_4(a):_{\HH}.
 \label{eq:h-mmt2-residual}
\end{equation}

\subsubsection{FPUT-\texorpdfstring{$\beta$}{beta}}

The $N+1$ bonds of Eq.~\eqref{eq:fput-bonds} obey the exact constraint
$\sum_{j=0}^{N}r_j=0$.  The finite-size Hartree reference must preserve this
constraint:
\begin{equation}
 \mathcal H_{\HH,q}=\frac{\kappa_{\HH}}2\sum_{j=0}^{N}r_j^2,
 \qquad
 \langle r_i r_j\rangle_{\HH}=\frac{T}{\kappa_{\HH}}
 \left(\delta_{ij}-\frac{1}{N+1}\right).
 \label{eq:h-fput-covariance}
\end{equation}
Consequently,
\begin{align}
 \sigma_{\HH}^2&\equiv\langle r_j^2\rangle_{\HH}
 =\frac{TN}{(N+1)\kappa_{\HH}},
 &\kappa_{\HH}&=1+3b\sigma_{\HH}^2,
 \notag\\
 \kappa_{\HH}&=
 \frac{1+\sqrt{1+12bT\,N/(N+1)}}{2}.
 \label{eq:h-fput-selfconsistency}
\end{align}
For the normal modes defined in Eq.~\eqref{eq:fput-normal-transform},
\begin{equation}
 \langle Q_k^2\rangle_{\HH}
 =\frac{T}{\kappa_{\HH}[\omega_k^{(0)}]^2},
 \qquad
 \Delta\mathcal H_{\HH}
 =\frac b4\sum_{j=0}^{N}:r_j^4:_{\HH},
 \label{eq:h-fput-reference}
\end{equation}
where
$:r_j^4:_{\HH}=r_j^4-6\sigma_{\HH}^2r_j^2+3\sigma_{\HH}^4$.
The factors $N/(N+1)$ in Eqs.~\eqref{eq:h-fput-covariance} and
\eqref{eq:h-fput-selfconsistency} are retained throughout.

\subsubsection{DNLS}

For the finite ring of Eq.~\eqref{eq:dnls-hamiltonian}, the Hartree reference
to $\mathcal H_G=H-\mu\mathcal N$ is
\begin{align}
 \mathcal H_{\HH}&=\sum_k\xi_k^{\HH}|a_k|^2,
 &\xi_k^{\HH}&=\omega_k^{(0)}-\mu+2b\nu_{\HH},
 \notag\\
 n_k^{\HH}&=\frac{T}{\xi_k^{\HH}},
 &\nu_{\HH}&=\frac1L\sum_kn_k^{\HH}.
 \label{eq:h-dnls-reference}
\end{align}
Here $\nu_{\HH}$ is the density of the reference Gaussian itself.  Using a
density taken from the interacting system would leave an uncancelled
quadratic term and would not define a stationary Hartree expansion.  The
identity
\begin{equation}
 |\psi_j|^4=:|\psi_j|^4:_{\HH}
 +4\nu_{\HH}|\psi_j|^2-2\nu_{\HH}^2
 \label{eq:h-dnls-normal-ordering}
\end{equation}
then gives
\begin{equation}
 \Delta\mathcal H_{\HH}=\frac b2\sum_{j=0}^{L-1}
 :|\psi_j|^4:_{\HH}.
 \label{eq:h-dnls-residual}
\end{equation}

\subsection{Connected expansion and the vanishing first-order correction}
\label{subsec:hartree-connected}

Introduce a bookkeeping parameter multiplying the complete residual,
\begin{equation}
 \mathcal H_{\epsilon}=\mathcal H_{\HH}
 +\epsilon\Delta\mathcal H_{\HH}.
 \label{eq:h-interpolation}
\end{equation}
The self-consistent reference is held fixed as $\epsilon$ varies;
$\epsilon=0$ gives the Gaussian measure and $\epsilon=1$ recovers the
physical Gibbs measure, up to the irrelevant scalar in
Eq.~\eqref{eq:h-decomposition}.  Let
\begin{equation}
 I_k=\begin{cases}
 |a_k|^2,&\text{MMT and DNLS},\\
 Q_k^2,&\text{FPUT}.
 \end{cases}
 \label{eq:h-observable}
\end{equation}
Reweighting the Gaussian measure gives the exact identity
\begin{equation}
 \langle I_k\rangle_{\epsilon}=
 \frac{\left\langle I_k
 \exp(-\beta\epsilon\Delta\mathcal H_{\HH})\right\rangle_{\HH}}
 {\left\langle
 \exp(-\beta\epsilon\Delta\mathcal H_{\HH})\right\rangle_{\HH}}.
 \label{eq:h-reweighting}
\end{equation}
Its formal expansion in connected cumulants is \cite{Kubo1962}
\begin{align}
 \frac{\langle I_k\rangle_{\epsilon}}{n_k^{\HH}}
 &=1+\sum_{m=1}^{\infty}a_{m,k}\epsilon^m,
 \label{eq:h-series}\\
 a_{m,k}&=\frac{(-\beta)^m}{m!\,n_k^{\HH}}
 \left\langle I_k;
 \underbrace{\Delta\mathcal H_{\HH};\ldots;
 \Delta\mathcal H_{\HH}}_{m\ \mathrm{copies}}
 \right\rangle_{\HH,c}.
 \label{eq:h-cumulants}
\end{align}
For FPUT, $n_k^{\HH}$ in these two equations denotes
$\langle Q_k^2\rangle_{\HH}$.

At first order,
\begin{equation}
 a_{1,k}=-\frac{\beta}{n_k^{\HH}}
 \operatorname{Cov}_{\HH}(I_k,\Delta\mathcal H_{\HH}).
 \label{eq:h-a1-covariance}
\end{equation}
Write $I_k=n_k^{\HH}+:I_k:_{\HH}$.  The centered part of the observable has
Gaussian degree two, whereas every term in the residual is a fully
normal-ordered polynomial of degree three or four.  Orthogonality of Gaussian
Wick polynomials therefore gives
\begin{equation}
 \langle\Delta\mathcal H_{\HH}\rangle_{\HH}=0,
 \qquad
 \langle:I_k:_{\HH}\Delta\mathcal H_{\HH}\rangle_{\HH}=0,
 \qquad a_{1,k}=0.
 \label{eq:h-a1-zero}
\end{equation}
This argument also applies to the correlated FPUT covariance in
Eq.~\eqref{eq:h-fput-covariance}, provided that normal ordering is performed
with the full constrained covariance.  Thus the cancellation is a consequence
of self-consistency and Gaussian-degree orthogonality, not of phase symmetry
alone.

\subsection{Leading nonzero correction}
\label{subsec:hartree-a2}

The first nonzero correction is the second-order coefficient
\begin{align}
 a_{2,k}&=\frac{\beta^2}{2n_k^{\HH}}
 \langle I_k;\Delta\mathcal H_{\HH};
 \Delta\mathcal H_{\HH}\rangle_{\HH,c}
 \notag\\
 &=\frac{\beta^2}{2n_k^{\HH}}
 \left[\langle I_k(\Delta\mathcal H_{\HH})^2\rangle_{\HH}
 -n_k^{\HH}\langle(\Delta\mathcal H_{\HH})^2\rangle_{\HH}\right].
 \label{eq:h-a2-general}
\end{align}
The complete residual of Eq.~\eqref{eq:h-decomposition} must be retained in
this expression.

For MMT branch I, exact complex-Gaussian contraction gives
\begin{equation}
 a_{2,k}^{(I)}=2\mathcal R_{M,k}h^2,
 \qquad
 \mathcal R_{M,k}=\sum_{i,j,r=1}^{M}
 \mathbf 1_{\{i+j=r+k\}},
 \qquad h=\frac{bT}{A^2}.
 \label{eq:h-mmt1-a2}
\end{equation}
For $M=8$, the integer factors $2\mathcal R_{8,k}$ are
\begin{equation}
 (72,84,92,96,96,92,84,72).
 \label{eq:h-mmt1-routing}
\end{equation}
For branch II, define
\begin{equation}
 \mathcal S_k=\sum_{i,j,r=1}^{M}
 n_i^{\HH}n_j^{\HH}n_r^{\HH}
 \mathbf 1_{\{i+j=r+k\}}.
 \label{eq:h-mmt2-sum}
\end{equation}
The mixed cubic--quartic contraction vanishes by odd Gaussian degree, and
\begin{align}
 a_{2,k}^{(3)}={}&\beta^2b n_k^{\HH}
 \left[4\sum_{j=1}^{M-k}n_j^{\HH}n_{j+k}^{\HH}
 +2\sum_{i=1}^{k-1}n_i^{\HH}n_{k-i}^{\HH}\right],
 \notag\\
 a_{2,k}^{(4)}={}&2\beta^2b^2n_k^{\HH}\mathcal S_k,
 \qquad
 a_{2,k}^{(II)}=a_{2,k}^{(3)}+a_{2,k}^{(4)}.
 \label{eq:h-mmt2-a2}
\end{align}

For the constrained FPUT chain, the result is independent of $k$:
\begin{equation}
 a_{2,k}^{\mathrm{FPUT}}=
 6\frac{N^2-N+1}{(N+1)^2}h^2,
 \qquad h=\frac{bT}{\kappa_{\HH}^2}.
 \label{eq:h-fput-a2}
\end{equation}
The finite-size factor follows from carrying out the contractions with
Eq.~\eqref{eq:h-fput-covariance}; replacing the constrained bonds by
independent variables would miss this factor.  For DNLS, with all indices
understood modulo $L$, let
\begin{equation}
 \mathcal T_k=\sum_{p,q=0}^{L-1}
 n_p^{\HH}n_q^{\HH}n_{p+q-k}^{\HH}.
 \label{eq:h-dnls-sunset}
\end{equation}
The complete normal-ordered quartet gives
\begin{equation}
 a_{2,k}^{\mathrm{DNLS}}=
 \frac{2\beta^2b^2}{L^2}n_k^{\HH}\mathcal T_k.
 \label{eq:h-dnls-a2}
\end{equation}
The closed contractions were also checked by direct Gaussian polynomial
enumeration.  Details of the contraction tables are given in the Supplementary
Material.

\subsection{Finite-order comparison}
\label{subsec:hartree-comparison}

\begin{figure}[!htbp]
 \centering
 \includegraphics[width=0.98\textwidth]{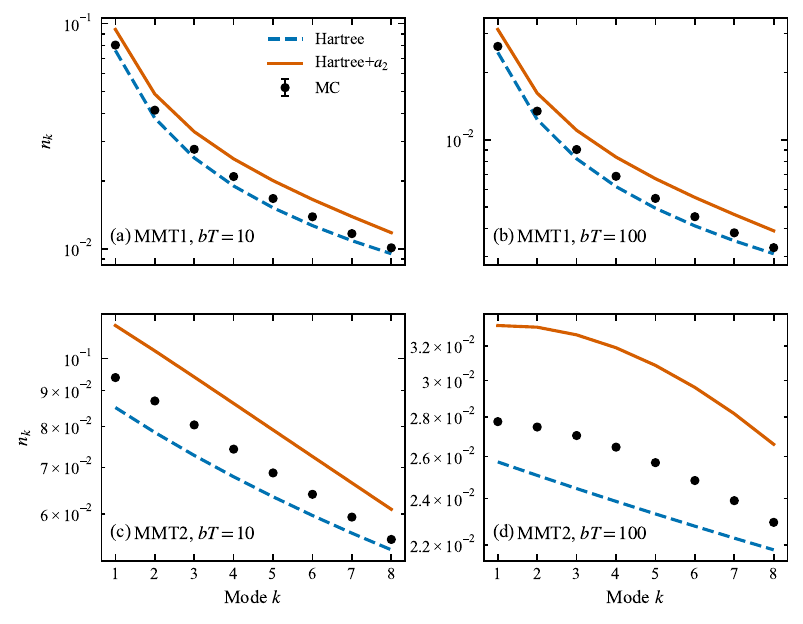}
 \caption{Self-consistent Gaussian/Hartree results for the two MMT branches.
 Panels (a,b) show branch I and panels (c,d) branch II for $M=8$, $T=1$, and
 $bT=10,100$.  Symbols are independent MC data, dashed curves are
 $n_k^{\HH}$, and solid curves are $n_k^{\HH}(1+a_{2,k})$.}
 \label{fig:h-mmt}
\end{figure}

\begin{figure}[!htbp]
 \centering
 \includegraphics[width=0.98\textwidth]{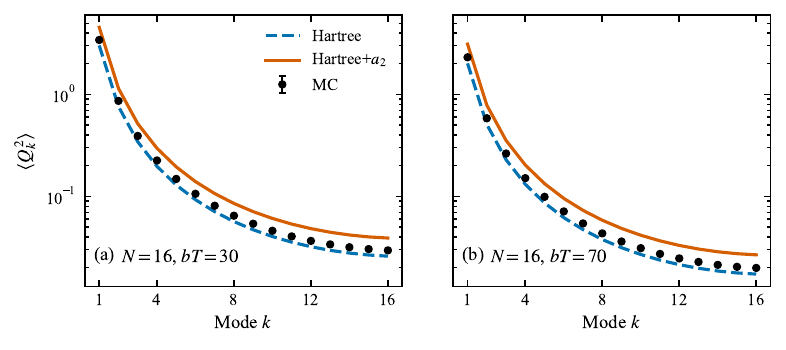}
 \caption{FPUT-$\beta$ modal variance for $N=16$, $T=1$: (a) $bT=30$ and
 (b) $bT=70$.  Symbols and curves have the same meaning as in
 Fig.~\ref{fig:h-mmt}.  Both the Hartree stiffness and $a_2$ retain the exact
 finite-size constraint factor.}
 \label{fig:h-fput}
\end{figure}

\begin{figure}[!htbp]
 \centering
 \includegraphics[width=0.98\textwidth]{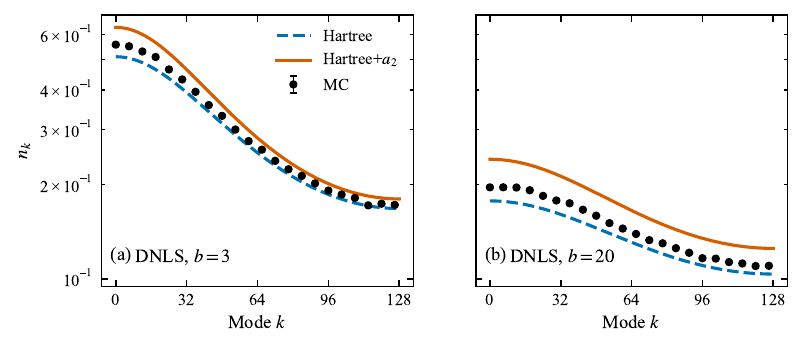}
 \caption{DNLS occupations in the normal regime for $L=256$, $T=1$, and
 $\mu=-0.2$: (a) $b=3$ and (b) $b=20$. Symbols are independent MC data. The dashed curves use the
 self-consistent reference density in Eq.~\eqref{eq:h-dnls-reference}; the
 solid curves include the complete second-order contribution in
 Eq.~\eqref{eq:h-dnls-a2}.}
 \label{fig:h-dnls}
\end{figure}

Figures~\ref{fig:h-mmt}--\ref{fig:h-dnls} compare independent MC data with
the Hartree result and the second-order partial sum
\begin{equation}
 n_k^{(2)}=n_k^{\HH}(1+a_{2,k}).
 \label{eq:h-second-order}
\end{equation}
The parameters and finite system sizes are those used for the nonperturbative
tests in Sec.~\ref{sec:nonperturbative}.  The latter results coincide with the
MC data at the resolution of the figures and are therefore not repeated as an
additional curve.  Quantitative errors are measured against the finite-size
nonperturbative result using
\begin{equation}
 \mathcal E_{\mathrm{MAR}}=
 \frac{1}{N_m}\sum_k
 \left|\frac{n_k^{\mathrm{approx}}-n_k^{\mathrm{NP}}}
 {n_k^{\mathrm{NP}}}\right|.
 \label{eq:h-mar-error}
\end{equation}

For the two MMT branches at $M=8$, the Hartree errors at $bT=10$ and $100$
are $7.7\%$ and $8.2\%$ for branch I, and $7.5\%$ and $8.1\%$ for branch
II.  The second-order coefficient is positive for every displayed mode and
moves the spectrum past the full result, increasing the corresponding errors
to $18.7\%$, $21.1\%$, $15.0\%$, and $19.4\%$.  The self-consistent
reference captures the modal dependence, but its residual interaction is not
small enough for the first nonzero correction to improve the approximation.

For FPUT-$\beta$, bond exchangeability makes $a_{2,k}$ independent of $k$,
so the correction changes the amplitude but not the modal dependence.  At
$N=16$ and $bT=30,70$, the Hartree errors are $12.4\%$ and $13.0\%$,
whereas the second-order errors are $31.8\%$ and $34.4\%$.  The finite-size
factor in Eq.~\eqref{eq:h-fput-a2} is used in both cases.

For DNLS in the normal regime, with $L=256$, $T=1$, and $\mu=-0.2$, the
Hartree error changes from $4.8\%$ to $7.9\%$ after including $a_2$ at
$b=3$, and from $7.8\%$ to $17.2\%$ at $b=20$.  Thus the complete
momentum-conserving contraction has the same overcorrection tendency as in
the other models.

The Hartree reference captures much of the nonlinear frequency shift and
gives a reasonable occupation profile. The remaining interaction is not
necessarily small, however. In these examples the first nonzero correction
already moves the result farther from the exact occupation. We therefore
examine the later terms rather than assuming that they will restore accuracy.

\subsection{Growth of higher-order corrections}
\label{subsec:hartree-high-order}

We use the actual power of $\epsilon$ to label the order and define
\begin{equation}
 S_m=1+\sum_{j=1}^{m}a_{j,k}.
 \label{eq:h-partial-sum}
\end{equation}
Table~\ref{tab:h-high-order} lists three representative sequences evaluated
through fifth order.  The MMT result is obtained by exact Wick contractions for the finite set of modes.  For FPUT, all coefficients are evaluated at the same finite
size $N=16$ used in Fig.~\ref{fig:h-fput}; the constrained Gaussian moments
and an independent Fourier representation of
$\delta(\sum_{j=0}^{N}r_j)$ give identical coefficients.  The DNLS sequence
is obtained from the connected contractions through second order and from the
finite-ring transfer-operator interpolation at higher orders.

\begin{table}[t]
 \centering
 \caption{Coefficients of the fixed-reference Hartree expansion at selected
 strong-nonlinearity parameters.  The FPUT sequence is a finite-$N$ result.}
 \label{tab:h-high-order}
 \begin{tabular}{lrrrrr}
  \toprule
  system & $a_1$ & $a_2$ & $a_3$ & $a_4$ & $a_5$\\
  \midrule
  MMT I, $M=8$, $bT=10$, $k=1$
   & $0$ & $0.240$ & $-0.658$ & $2.05$ & $-7.68$\\
  FPUT-$\beta$, $N=16$, $bT=30$
   & $0$ & $0.505$ & $-1.73$ & $7.98$ & $-45.5$\\
  DNLS, $L=256$, $\mu=-0.2$, $b=20$, $k=1$
   & $0$ & $0.357$ & $-0.934$ & $3.15$ & $-12.9$\\
  \bottomrule
 \end{tabular}
\end{table}

For the finite FPUT chain, the Hartree stiffness and complete nonlinear
amplitude are
\begin{equation}
 \kappa_{\HH}=9.7172,
 \qquad
 \frac{T\theta_{16}}{T/\kappa_{\HH}}=\kappa_{\HH}\theta_{16}=1.1418,
 \label{eq:h-fput-finite-target}
\end{equation}
and the successive partial sums are
\begin{equation}
 (S_1,S_2,S_3,S_4,S_5)
 =(1,\ 1.505,\ -0.228,\ 7.754,\ -37.77).
 \label{eq:h-fput-finite-partial-sums}
\end{equation}
Thus the finite-size calculation has already departed from the physical
value after the second-order overshoot and then oscillates with rapidly
increasing amplitude.

\begin{figure}[!htbp]
 \centering
 \includegraphics[width=0.98\textwidth]{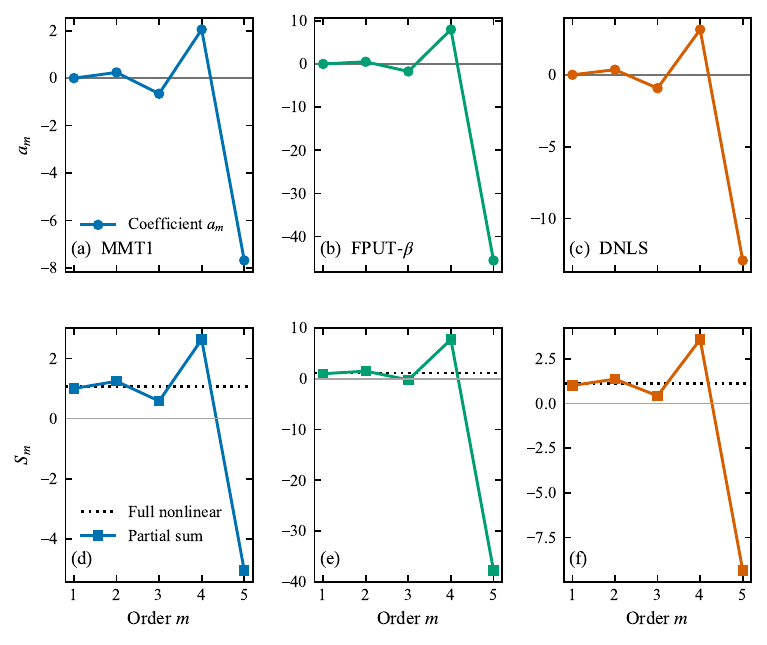}
 \caption{Loss of perturbative ordering about the fixed Hartree reference.
 The upper panels show $a_m$ at actual orders $m=1,\ldots,5$ for (a) MMT
 branch I, $M=8$, $bT=10$, $k=1$; (b) FPUT-$\beta$, $N=16$, $bT=30$; and
 (c) DNLS, $L=256$, $T=1$, $\mu=-0.2$, $b=20$, $k=1$.  The lower panels
 show the corresponding partial sums.  Dotted lines are the normalized full
 nonlinear results at the same finite sizes.}
 \label{fig:h-high-order}
\end{figure}

The three sequences in Fig.~\ref{fig:h-high-order} exhibit the same loss of
ordering.  After the positive $a_2$ term, the signs alternate and the
magnitudes increase rapidly.  By fifth order, $|a_5|/|a_2|$ is approximately
$32$, $90$, and $36$ for MMT, FPUT, and DNLS, respectively.  Increasing the
order of the same fixed-reference expansion therefore does not generate a
sequence of improving approximations at these parameters.

The coefficients in Table~\ref{tab:h-high-order} quantify the failure
through the calculated fifth order, consistent with the large-order behavior
of anharmonic expansions \cite{BenderWu1969,BenderWu1971}. The mixed MMT
branch is tested at second order in Fig.~\ref{fig:h-mmt}.

The underlying scale balance is already visible in the reference problem.
For MMT branch I, $h=bT/A^2$ tends to $1/(2M)$ as $bT\to\infty$;
for FPUT, $h=bT/\kappa_{\HH}^2$ tends to $(N+1)/(3N)$.
Dressing the quadratic scale therefore does not make the residual vanish
in the strong-nonlinearity limit. Mode sums and connected contractions
multiply these finite factors, producing the growing corrections in
Fig.~\ref{fig:h-high-order}. The useful Hartree spectrum and the subsequent
failure of its residual expansion have the same origin: the reference
captures a mean interaction scale but not the full fluctuations around it.

A second reorganization absorbs the mean diagonal four-wave processes into
dressed modal scales and removes the associated trivial-pairing families from
the explicit residual.  Such renormalized-wave constructions have been used
for nonlinear lattices \cite{Gershgorin2005,Gershgorin2007}.  The next section
treats this construction separately and examines
whether its improved low-order behavior is accompanied by controlled
higher-order corrections.

\section{Perturbation Theory with Trivial-Pairing Renormalization}
\label{sec:projected-closure}

The Hartree calculation retains all fluctuations left after the mean
interaction has been absorbed. A second construction treats the diagonal
four-wave processes more selectively. These processes pair each incoming
mode with an outgoing mode of the same index and are often called trivial
pairings. Their mean contribution changes the modal frequency or stiffness
\cite{Gershgorin2005,Gershgorin2007}.

Here we determine that shift from the corrected occupations and construct
the explicit perturbative terms from the remaining, non-pairing
interactions. This feedback can give accurate low-order occupations at
strong nonlinearity. We first specify which fluctuations are omitted,
then test the low-order result and the growth of later corrections.

\subsection{Diagonal and non-pairing interactions}
\label{sec:projected-interaction}

For MMT and DNLS, write the quartic form as
\begin{equation}
 Q_4=\sum_{k_1,k_2,k_3,k_4}^{\Gamma}
 W_{1234}\,a_{k_1}a_{k_2}a_{k_3}^{*}a_{k_4}^{*},
 \qquad
 W_{1234}=w_{k_1}w_{k_2}w_{k_3}w_{k_4},
 \label{eq:projected-generic-quartet}
\end{equation}
where $\Gamma$ is the model-specific mode-selection rule.  For MMT,
$k_1+k_2=k_3+k_4$ and $w_k=k^{c/4}$; for DNLS the equality is understood
modulo $L$ and $w_k=1$.  The union of the two complex Gaussian pairings is
selected without double counting by
\begin{align}
 \chi_D(1,2;3,4)
 &=\delta_{13}\delta_{24}+\delta_{14}\delta_{23}
   -\delta_{13}\delta_{14}\delta_{23}\delta_{24},
 \notag\\
 \chi_{**}(1,2;3,4)
 &=(1-\delta_{13}\delta_{24})(1-\delta_{14}\delta_{23})
 =1-\chi_D .
 \label{eq:projected-characters}
\end{align}
The final term in $\chi_D$ removes the double counting when all four
indices coincide.

With
\begin{equation}
 u_k=w_k^2,\qquad I_k=|a_k|^2,\qquad
 S=\sum_k u_k I_k,
 \label{eq:projected-weighted-intensity}
\end{equation}
the diagonal and complementary quartic forms are
\begin{equation}
 Q_D=2S^2-\sum_k u_k^2 I_k^2,
 \qquad Q_{**}=Q_4-Q_D .
 \label{eq:projected-quartet-union}
\end{equation}
Write $V_4=(g_4/2)Q_4$, where $g_4=b$ for MMT and $g_4=b/L$ for
DNLS with the unitary Fourier convention of Eq.~\eqref{eq:dnls-fourier}.
Let $S_*$ be a prescribed moment entering the quadratic reference,
\begin{equation}
 K_{\mathrm R}=K_0+2g_4S_*S,\qquad
 n_k^{\mathrm R}=\frac{T}{\varepsilon_k+2g_4S_*u_k}.
 \label{eq:projected-reference}
\end{equation}
Here $K$ denotes the Gibbs generator, $K_0=\sum_k\varepsilon_k I_k$,
with $\varepsilon_k=k$ for MMT and
$\varepsilon_k=\omega_k^{(0)}-\mu$ for DNLS.
The moment $S_*$ is distinct from the Gaussian expectation
$\langle S\rangle_{\mathrm R}=\sum_k u_k n_k^{\mathrm R}$ unless a
Hartree self-consistency condition is imposed. Adding and subtracting the
same quadratic term gives
\begin{align}
 K&=K_{\mathrm R}+H_3+V_{**}+D_*+\mathrm{const},
 \label{eq:projected-exact-decomposition}\\
 V_{**}&=\frac{g_4}{2}Q_{**},\qquad
 D_*=g_4(S-S_*)^2-\frac{g_4}{2}\sum_k u_k^2I_k^2 .
 \label{eq:projected-diagonal-fluctuation}
\end{align}
The additive constant is $-g_4S_*^2$ and cancels from normalized averages.
$H_3$ is present only in MMT branch II. Retaining $D_*$ preserves the
fluctuations of the diagonal intensities; the projected closure keeps
$H_3+V_{**}$ and omits $D_*$. This is the approximation that distinguishes it
from the full Hartree residual expansion. In DNLS, $u_k=1$ and
$S=\mathcal N$, so the shift is $2bS_*/L$.

For the real FPUT coordinates, the corresponding projector deletes quartic
monomials with multiplicity patterns $4$ and $2+2$, the union of the three
real Gaussian pairings. The complex and real projectors are derived in the
Supplementary Material.

The projected polynomial $V_{**}$ is generally sign-indefinite.  Although all
of its finite Gaussian moments exist, $K_{\mathrm R}+\epsilon V_{**}$ need not
define a normalizable Gibbs measure for $\epsilon>0$.  Hence $\epsilon$ is
only a formal order marker, and the projected coefficients must not be
interpreted as Taylor coefficients of an exact Gibbs interpolation whose
$\epsilon=1$ endpoint is the physical Hamiltonian.

For $O_k=|a_k|^2$ in a complex model and $O_k=Q_k^2$ in FPUT, define the
connected Gaussian coefficients \cite{Kubo1962}
\begin{equation}
 a_{m,k}^{**}=
 \frac{(-\beta)^m}{m!\langle O_k\rangle_{\mathrm R}}
 \left\langle
 O_k;\underbrace{V_{\mathrm R,**};\ldots;V_{\mathrm R,**}}_{m\ \mathrm{times}}
 \right\rangle_{\mathrm R,c},
 \qquad
 \frac{\langle O_k\rangle^{[m]}}{\langle O_k\rangle_{\mathrm R}}
 =1+\sum_{j=1}^{m}a_{j,k}^{**}.
 \label{eq:projected-cumulants}
\end{equation}
For MMT branch II,
$V_{\mathrm R,**}=H_3+(g_4/2)Q_{**}$; in the other models it is the projected
quartic residual.  The first-order coefficient is
\begin{equation}
 a_{1,k}^{**}=-\frac{\beta}{\langle O_k\rangle_{\mathrm R}}
 \langle O_k;V_{\mathrm R,**}\rangle_{\mathrm R,c}=0.
 \label{eq:projected-a1-zero}
\end{equation}
For a complex quartet, every nonzero contraction of $O_kV_4$ selects one of
the two deleted pairings.  For FPUT, multiplication by $Q_k^2$ does not alter
the odd modal multiplicities retained by the real projector.  The MMT cubic
term also gives zero at first order by Gaussian parity.  The leading nonzero
coefficient is therefore
\begin{equation}
 a_{2,k}^{**}=
 \frac{\beta^2}{2\langle O_k\rangle_{\mathrm R}}
 \langle O_k;V_{\mathrm R,**};V_{\mathrm R,**}\rangle_{\mathrm R,c}.
 \label{eq:projected-a2-general}
\end{equation}

\subsection{Self-consistent second-order closure}
\label{sec:projected-second-order}

The numerical effect of Eq.~\eqref{eq:projected-a2-general} depends on how the
moment entering the dressed scale is determined.  Freezing the ordinary
Hartree scale before adding $a_{2,k}^{**}$ systematically overcorrects the
spectra considered below.  Supplying the exact nonlinear moment removes much
of that error, but it does not in general give an independent prediction.  We
therefore close the dressed scale with the corrected occupation itself:
\begin{equation}
 n_k^{[2]}=n_k^{\mathrm R}(\Theta_2)
 \left[1+a_{2,k}^{**}(\Theta_2)\right],
 \qquad
 \Theta_2=\mathcal F[\bm n^{[2]}],
 \label{eq:projected-a2-closed}
\end{equation}
where $\Theta_2$ is the model-dependent frequency shift or stiffness and
$\mathcal F$ is the mean-field functional generating the dressed quadratic
term. For the complex fields, this sets $S_*=\sum_k u_k n_k^{[2]}$,
not $\sum_k u_k n_k^{\mathrm R}$.  Equation~\eqref{eq:projected-a2-closed} is an order-dependent
resummation rather than the second partial sum of a fixed-reference series:
re-expanding its implicit solution about the bare problem generates diagonal
insertions at arbitrarily high powers of the bare interaction.

For MMT branch I, the closed equations at $M=8$ are
\begin{align}
 n_k^{\mathrm R}&=\frac{T}{A_2k},
 &A_2&=1+2b\sum_{p=1}^{M}p\,n_p^{[2]},
 \notag\\
 a_{2,k}^{**}&=r_k\left(\frac{bT}{A_2^2}\right)^2,
 &(r_1,\ldots,r_8)&=(42,54,62,66,66,62,54,42).
 \label{eq:projected-mmt1-closure}
\end{align}
For MMT branch II,
\begin{equation}
 n_k^{\mathrm R}=\frac{T}{k+\Delta_2},
 \qquad
 \Delta_2=2b\sum_{p=1}^{M}n_p^{[2]},
 \label{eq:projected-mmt2-reference}
\end{equation}
and the cubic and quartic contributions give
\begin{align}
 a_{2,k}^{**}={}&\beta^2b n_k^{\mathrm R}
 \left[
 4\sum_{j=1}^{M-k}n_j^{\mathrm R}n_{j+k}^{\mathrm R}
 +2\sum_{i=1}^{k-1}n_i^{\mathrm R}n_{k-i}^{\mathrm R}
 \right]
 \notag\\
 &+2\beta^2b^2 n_k^{\mathrm R}
 \left[
 \sum_{\substack{i,j,r=1\\i+j=r+k}}^{M}
 n_i^{\mathrm R}n_j^{\mathrm R}n_r^{\mathrm R}
 -2n_k^{\mathrm R}\sum_{p=1}^{M}(n_p^{\mathrm R})^2
 +(n_k^{\mathrm R})^3
 \right].
 \label{eq:projected-mmt2-a2}
\end{align}
The final term restores the overlap of the two deleted pairing families at
finite $M$.

For the fixed-boundary FPUT-$\beta$ chain,
\begin{align}
 \langle Q_k^2\rangle_{\mathrm R}
 &=\frac{T}{\kappa_2[\omega_k^{(0)}]^2},
 &\langle Q_k^2\rangle^{[2]}
 &=\langle Q_k^2\rangle_{\mathrm R}(1+a_2^{**}),
 \notag\\
 a_2^{**}
 &=\frac{6(N-2)^2}{(N+1)^2}
 \left(\frac{bT}{\kappa_2^2}\right)^2,
 &\kappa_2
 &=1+\frac{3b}{N+1}\sum_{p=1}^{N}
 [\omega_p^{(0)}]^2\langle Q_p^2\rangle^{[2]}.
 \label{eq:projected-fput-closure}
\end{align}
The finite-$N$ factor is retained rather than replaced by its thermodynamic
limit.

For a periodic DNLS ring, set
\begin{equation}
 n_k^{\mathrm R}=\frac{T}{\omega_k^{(0)}-\mu+2b\nu_2},
 \qquad
 \nu_2=\frac1L\sum_{p=0}^{L-1}n_p^{[2]},
 \label{eq:projected-dnls-reference}
\end{equation}
and define the momentum-conserving convolution with the diagonal pairings removed
\begin{align}
 \mathcal T_k^{**}
 & =\sum_{i,j,r=0}^{L-1}
 n_i^{\mathrm R}n_j^{\mathrm R}n_r^{\mathrm R}
 \delta_{i+j-r-k}^{(L)}
 -2n_k^{\mathrm R}\sum_p(n_p^{\mathrm R})^2
 +(n_k^{\mathrm R})^3,
 \notag\\
 a_{2,k}^{**}
 &=8\beta^2\left(\frac{b}{2L}\right)^2
 n_k^{\mathrm R}\mathcal T_k^{**}.
 \label{eq:projected-dnls-a2}
\end{align}
Equations~\eqref{eq:projected-a2-closed} and
\eqref{eq:projected-dnls-reference}--\eqref{eq:projected-dnls-a2} are solved
together with the exact finite-ring convolution.

\subsection{Occupations and connected correlations}
\label{sec:projected-results}

\begin{figure}[!htbp]
 \centering
 \includegraphics[width=0.96\textwidth]{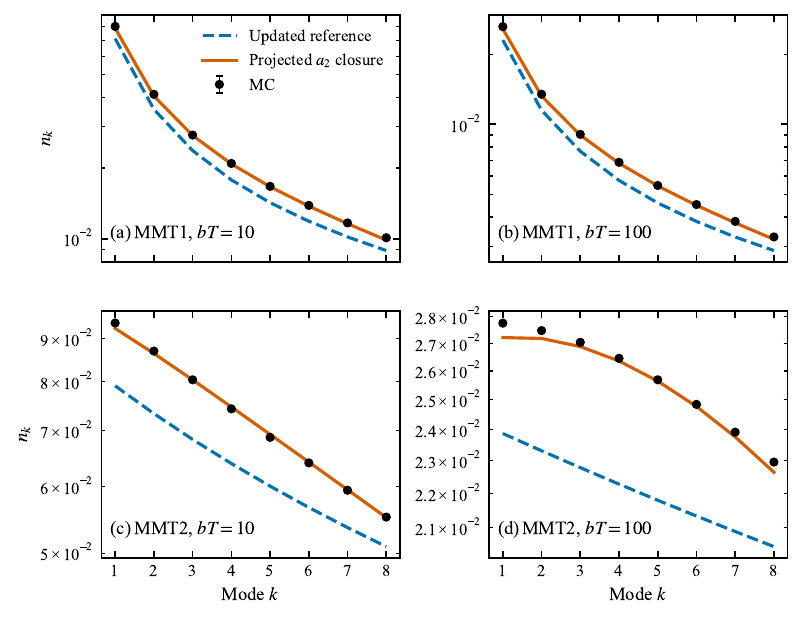}
 \caption{Trivial-pairing-renormalized low-order spectra for the two MMT
 branches at $M=8$ and $T=1$.  Panels (a,b) show branch I at $bT=10$ and
 $100$; panels (c,d) show branch II at the same nonlinearities.  Black points
 are MC means with replica standard errors.  The dashed blue curve is the
 updated Gaussian reference occupation, and the solid orange curve is the self-consistent
 projected $a_2$ closure.}
 \label{fig:projected-mmt}
\end{figure}

\begin{figure}[!htbp]
 \centering
 \includegraphics[width=0.96\textwidth]{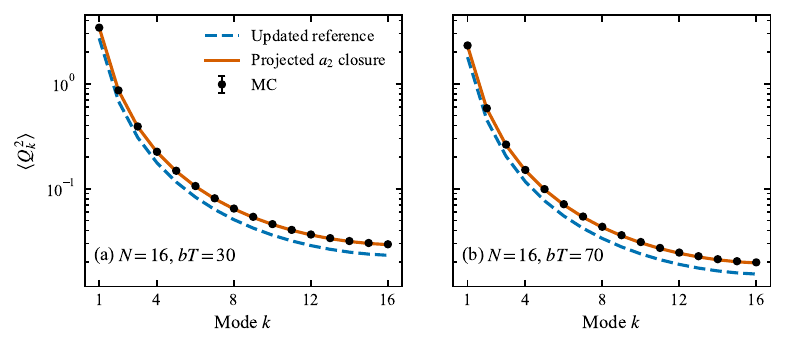}
 \caption{Fixed-boundary FPUT-$\beta$ modal variances for $N=16$, $T=1$,
 and (a) $bT=30$ and (b) $bT=70$.  The exact finite-$N$ projector and
 stiffness closure are used.  Symbols and curves have the same meaning as in
 Fig.~\ref{fig:projected-mmt}.}
 \label{fig:projected-fput}
\end{figure}

\begin{figure}[!htbp]
 \centering
 \includegraphics[width=0.96\textwidth]{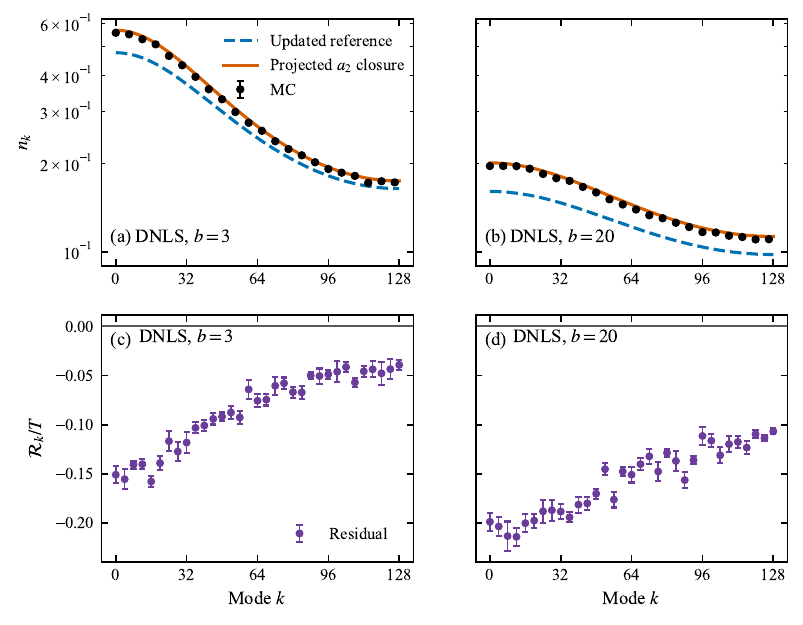}
 \caption{DNLS results in the normal regime for $L=256$, $T=1$,
 $\mu=-0.2$, and $b=3$ (left) or $20$ (right).  Panels (a,b) compare the
 modal occupations from MC, the updated Gaussian reference occupation, and the projected
 $a_2$ closure.  Panels (c,d) show the normalized residual
 $\mathcal R_k/T$ of the scalar Hartree/Rayleigh--Jeans relation for the same
 simulations.  All convolutions and density shifts retain the finite ring.}
 \label{fig:projected-dnls}
\end{figure}

Figures~\ref{fig:projected-mmt}--\ref{fig:projected-dnls} compare the
self-consistent $a_2$ closure with independent MC simulations.  In every
panel the dashed curve is the Gaussian reference occupation evaluated at the converged
$a_2$-closed scale, whereas the solid curve also includes the explicit factor
$1+a_{2,k}^{**}$.  The dashed curve identifies the reference contribution,
whereas the separation between the two curves shows the explicit projected
correction.

For DNLS, the accuracy of the spectrum does not imply that a scalar
Hartree/Rayleigh--Jeans relation has become exact.  Integration by parts in the full Gibbs
measure gives
\begin{align}
 T={}&[\omega_k^{(0)}-\mu]n_k
 +\frac bL\sum_{i,j}
 \left\langle a_k^*a_{i+j-k}^*a_i a_j\right\rangle,
 \notag\\
 \mathcal R_k={}&T-[\omega_k^{(0)}-\mu+2b \bar{n}]n_k
 =\frac bL\sum_{i,j}
 \left\langle a_k^*a_{i+j-k}^*a_i a_j\right\rangle_c .
 \label{eq:projected-dnls-residual}
\end{align}
Thus $\mathcal R_k=0$ would be required by exact Wick factorization with a
single scalar shift.  The lower panels of Fig.~\ref{fig:projected-dnls} use
the full finite-ring density and plot
$\mathcal R_k/T=1-[\omega_k^{(0)}-\mu+2b \bar{n}]n_k^{\mathrm{MC}}/T$.

Table~\ref{tab:projected-errors} compares the analytic approximations
with the full finite-system nonperturbative result.  The mean absolute relative error is the quantity
defined in Eq.~\eqref{eq:h-mar-error}.

\begin{table}[t]
 \centering
 \caption{Mean absolute relative error (percent) of the updated Gaussian
 factor and the self-consistent projected $a_2$ closure, measured against the
 complete finite-system nonperturbative result.}
 \label{tab:projected-errors}
 \begin{tabular}{lccc}
  \toprule
  Model & parameters & updated reference & projected $a_2$ closure\\
  \midrule
  MMT branch I  & $bT=10$  & 13.49 & 1.05\\
  MMT branch I  & $bT=100$ & 14.56 & 1.24\\
  MMT branch II & $bT=10$  & 12.78 & 0.64\\
  MMT branch II & $bT=100$ & 14.25 & 0.87\\
  FPUT-$\beta$ & $N=16,\ bT=30$ & 21.62 & 0.96\\
  FPUT-$\beta$ & $N=16,\ bT=70$ & 22.66 & 1.30\\
  DNLS & $L=256,\ b=3$  & 8.37 & 1.60\\
  DNLS & $L=256,\ b=20$ & 14.68 & 1.94\\
  \bottomrule
 \end{tabular}
\end{table}

The same construction therefore reduces the mean error to approximately
$0.6$--$2.0\%$ in three algebraically different classes of systems.  In MMT
branch I the $k^{-1}$ reference spectrum acquires finite-mode shape
corrections through the wave-number-dependent coefficients $r_k$;
branch II additionally tests the coexistence of cubic and quartic channels.
FPUT tests a constrained real field, and DNLS a dispersive complex field at
finite chemical potential.  The tests cover distinct interaction and constraint structures at
$bT=100$, $bT=70$, and $b=20$.

The mechanism is nevertheless selective.  The mean diagonal quartet has
already been absorbed into the dressed quadratic scale, and the feedback in
Eq.~\eqref{eq:projected-a2-closed} resums further diagonal insertions.  It
does not restore Gaussian statistics.  Indeed, the maximum absolute residual
in the lower panels of Fig.~\ref{fig:projected-dnls} is $0.170$ at $b=3$ and
$0.224$ at $b=20$.  The low-order spectral accuracy is therefore produced by
a compensating partial resummation, not by exact Wick factorization of the
nonlinear equilibrium state.

\subsection{Growth of corrections beyond low order}
\label{sec:projected-high-order}

To determine whether the low-order success extends systematically, we first
solve Eq.~\eqref{eq:projected-a2-closed} and then freeze the converged Gaussian
covariance.  At that fixed reference we enumerate the projected Gaussian
coefficients through formal order $m=5$.  The tests use MMT branch I
$(M=8,bT=10,k=1)$, FPUT-$\beta$ $(N=8,bT=30,k=1)$, and DNLS
$(L=5,T=1,\mu=-0.2,b=20,k=1)$.  Exact polynomial enumeration grows rapidly
with system size, so the smaller FPUT and DNLS systems are used only as
small-system tests of the higher-order coefficients. The low-order spectra
above use $N=16$ and $L=256$.

The resulting signed coefficients are
\begin{align}
 (a_1^{**},\ldots,a_5^{**})_{\mathrm{MMT\ I}}
 &=(0,\ 0.1081,\ -0.08671,\ 0.2064,\ -0.4401),
 \notag\\
 (a_1^{**},\ldots,a_5^{**})_{\mathrm{FPUT}}
 &=(0,\ 0.2093,\ -0.1954,\ 0.6226,\ -1.608),
 \notag\\
 (a_1^{**},\ldots,a_5^{**})_{\mathrm{DNLS}}
 &=(0,\ 0.1400,\ -0.1189,\ 0.2928,\ -0.6405).
 \label{eq:projected-high-order-coefficients}
\end{align}

\begin{figure}[!htbp]
 \centering
 \includegraphics[width=0.98\textwidth]{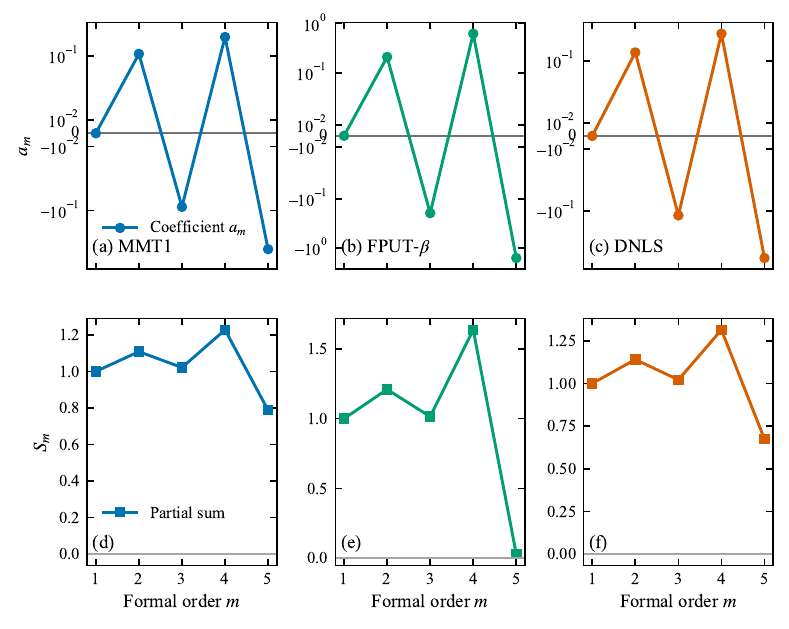}
 \caption{Higher-order coefficients about the fixed, second-order-closed
 reference. Panels (a--c) show the signed coefficients through order five for
 MMT branch I ($M=8$, $bT=10$, $k=1$), FPUT-$\beta$ ($N=8$, $bT=30$,
 $k=1$), and DNLS ($L=5$, $T=1$, $\mu=-0.2$, $b=20$, $k=1$).
 Panels (d--f) show the corresponding partial sums of the projected expansion.}

\label{fig:projected-high-order}
\end{figure}

In all three examples the coefficient decreases in magnitude from second
to third order, but the fourth and fifth terms then grow. The partial sums
oscillate with increasing amplitude. Thus the accurate low-order closure
does not lead to successively better approximations when further terms
are added. Its success comes from including selected interaction effects
through self-consistency, not from making all remaining fluctuations small.

\section{Infrared Failure of Perturbation Theory near DNLS Quasicondensation}
\label{sec:dnls-infrared-failure}

The nonperturbative results in Sec.~\ref{sec:dnls-quasicondensation}
remain accurate when the zero mode becomes highly populated.
We now return to this weak-coupling regime with the two perturbative
approximations defined in Secs.~\ref{sec:hartree}
and \ref{sec:projected-closure}. This comparison tests a different source
of failure from large bare nonlinearity: the growth of low-frequency
fluctuations and spatial coherence.

\subsection{Comparison of the two second-order approximations}
\label{subsec:dnls-ir-perturbative-failure}

\begin{figure}[!htbp]
 \centering
 \includegraphics[width=0.98\linewidth]{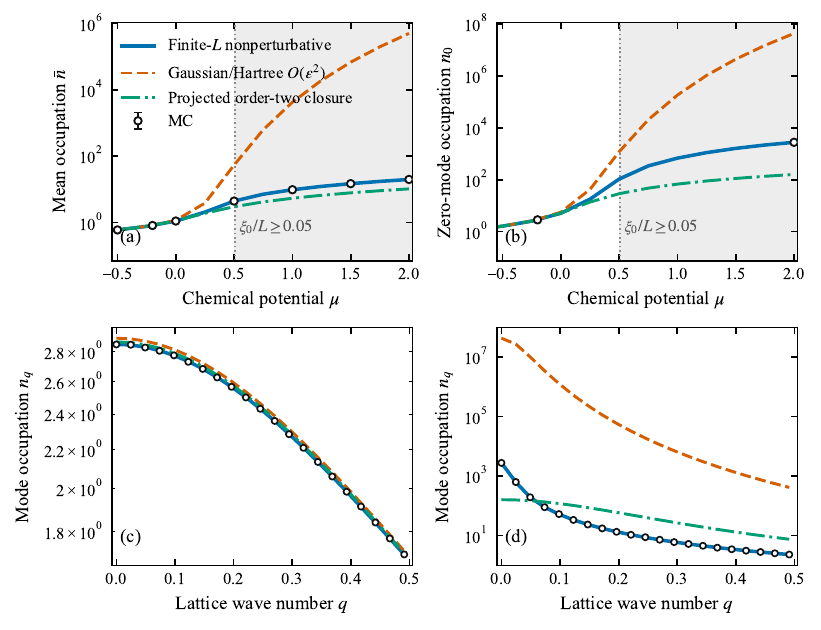}
 \caption{Finite-ring DNLS results across the quasicondensation crossover at
 $T=1$, $L=256$, and $b=0.1$.  (a) Mean norm density $\bar n$ and
 (b) zero-mode occupation $n_0$ versus chemical potential.  The shaded region
 begins where the exact diagnostic $\xi_0/L$ reaches $0.05$; this is a guide
 to the crossover, not a phase boundary.  (c) Modal occupations at the normal
 point $\mu=-0.2$.  (d) Modal occupations at the quasicondensed point $\mu=2$.
 Black symbols are independent MC results and error bars are replica standard
 errors.  The solid blue curve is the complete finite-ring nonperturbative
 prediction; the dashed orange and dash-dotted green curves are the Hartree
 second-order partial sum and the projected order-two closure, respectively.}
 \label{fig:dnls-ir-breakdown}
\end{figure}

The Hartree curve in Fig.~\ref{fig:dnls-ir-breakdown} is the strict
second-order partial sum of Eqs.~\eqref{eq:h-dnls-reference},
\eqref{eq:h-dnls-a2}, and \eqref{eq:h-second-order}.  The projected curve is
the self-consistent order-two closure of
Eqs.~\eqref{eq:projected-a2-closed},
\eqref{eq:projected-dnls-reference}, and
\eqref{eq:projected-dnls-a2}.  The Hartree first-order term vanishes by its self-consistent normal
ordering; the projected first-order term vanishes by the pairing projector.  The former is a
fixed-reference partial sum, while the latter feeds the projected correction
back into the density that fixes its reference shift.

At the normal point $\mu=-0.2$, both constructions follow the broad MC
spectrum reasonably well.  At $\mu=2$, however, both separate strongly from
the simulation in the lowest modes while the nonperturbative curve remains
accurate.  The nonperturbative and MC values were given in
Eq.~\eqref{eq:dnls-ir-np-mc-numbers}. The two approximations instead give
\begin{equation}
 (\bar n^{\mathrm{H2}},n_0^{\mathrm{H2}})
 =(5.06\times10^5,\ 4.32\times10^7),
 \qquad
 (\bar n^{\mathrm{TP2}},n_0^{\mathrm{TP2}})
 =(10.29,\ 159.9).
 \label{eq:dnls-ir-perturbative-numbers}
\end{equation}
Here H2 denotes the strict second-order Hartree partial sum, whereas TP2
denotes the self-consistent trivial-pairing-projected order-two closure. For H2 the self-consistent reference itself remains on its physical branch:
$r_{\mathrm H}=-\mu+2b\nu_{\mathrm H}=2.49\times10^{-3}>0$.
Its very large second-order value is instead generated by the infrared
convolution in Eq.~\eqref{eq:h-dnls-a2}.  The projected closure underestimates the infrared occupation. Both
constructions therefore fail to describe the crossover despite their
reasonable performance in the normal regime.

\subsection{Enhancement of low-frequency fluctuations}
\label{subsec:dnls-infrared-enhancement}

This failure cannot be diagnosed from $b$ alone.  At positive $\mu$, the
$b=0$ grand-canonical measure is non-normalizable because the quadratic
generator is unstable in the lowest modes.  The small-$b$ quasicondensed
regime is consequently not a regular perturbation of a well-defined
$b=0$ Gibbs state.  More generally, either dressed reference has the
long-wavelength form
\begin{equation}
 n_q^{\mathrm R}=\frac{T}{\omega^{(0)}(q)+r},
 \qquad
 \omega^{(0)}(q)=q^2+O(q^4),
 \qquad
 \ell_r=r^{-1/2}.
 \label{eq:dnls-ir-massive-reference}
\end{equation}
Products of such propagators generate sums
\begin{equation}
 I_p(r,L)=\frac1L\sum_k[\omega_k^{(0)}+r]^{-p}
 \simeq \ell_r^{2p-1}\Phi_p(L/\ell_r),
 \label{eq:dnls-ir-finite-size-scaling}
\end{equation}
with $\Phi_p(x)$ approaching a constant for $x\gg1$ and
$\Phi_p(x)\sim x^{-1}$ for $x\ll1$.  Hence
$I_p\sim\ell_r^{2p-1}$ before the correlation length reaches the ring and
$I_p\sim\ell_r^{2p}/L$ in the zero-mode-dominated limit.  The reference
length $\ell_r$ need not equal the exact transfer length $\xi_0$ in
Eq.~\eqref{eq:dnls-ir-correlation-length}, but both grow as the infrared
sector softens.  Consequently, products and momentum-constrained convolutions of these
infrared sums can offset a nominal prefactor $b^s$, so the corresponding
contribution need not remain small even when $b\ll1$.  The derivation of
Eq.~\eqref{eq:dnls-ir-finite-size-scaling} and its limiting forms is given in
the Supplementary Material.

The comparison shows why weak bare coupling alone is not enough.
Near quasicondensation, the large low-mode occupations amplify the residual
interaction. Both second-order approximations then fail, while the exact
finite-ring result remains accurate.

\section{Conclusions and Discussion}
\label{sec:conclusions-discussion}

The equilibrium occupations of the MMT models, the FPUT-$\beta$ chain,
and the DNLS ring can be calculated without expanding in the nonlinear
coupling. The finite-size representations derived here retain the
interactions and constraints of each Hamiltonian. Independent MC
calculations confirm their accuracy from weak to strong nonlinearity.
The FPUT result illustrates why a simple modal dependence can survive
strong interactions: bond symmetry and the fixed-length constraint preserve
the inverse-square frequency dependence, while the nonlinear amplitude
is determined by the full bond statistics.

The DNLS solution also describes the growth of low-mode occupation and
long-range coherence at small nonlinear coefficients. It extends the
equilibrium information obtained from transfer methods
\cite{Rasmussen2000,Johansson2004,Nunnenkamp2007} to the occupation of
each mode on a finite ring. The positive RJ decomposition associates
each contribution with a correlation length. A single RJ relation
results when the contributing frequency offsets coincide, and is
approximately valid when one contribution dominates. Distinct offsets
produce curvature in the inverse occupation. Near quasicondensation,
a contribution with a small weight can still strongly populate the zero
mode if its correlation length is large. Retaining these contributions
accounts for the accurate finite-ring occupations across the crossover.

The same occupations determine the mean frequency of the complex MMT
and DNLS modes through $\bar\omega_k=\mu+T/n_k$. The dynamical spectra
show that this prediction remains accurate when interactions broaden
the spectrum or produce several peaks. In the mixed MMT branch,
the harmonic and first-renormalized frequencies lie near different
local maxima, while the nonperturbative result gives the mean over the
complete spectrum.

The perturbative comparisons clarify both the usefulness and the limits
of frequency renormalization \cite{Gershgorin2005,Gershgorin2007,Lee2009,Lee2013}.
The self-consistent Hartree state captures a substantial mean interaction
effect, but its second-order correction worsens the occupations in the
strongly nonlinear examples studied here. The trivial-pairing construction,
with the corrected occupations fed back into the mean shift, reduces the
mean errors to about $0.6$--$2.0\%$. This improvement includes selected
higher-order effects through self-consistency. It does not make the
remaining fluctuations small: the representative sequences through
fifth order develop growing, oscillatory partial sums. The DNLS
connected correlations likewise remain nonzero even when the low-order
occupations are accurate.

An additional important conclusion is that nonperturbative theory can
be necessary even when the bare nonlinear coefficient is small.
In the DNLS quasicondensation regime, the growth of low-mode populations
amplifies the interaction and both tested second-order approximations
fail. Strong microscopic interactions and weak-coupling collective
coherence therefore provide two distinct reasons to use the full
nonperturbative occupations. These results give a quantitative basis
for studying the distribution of optical power, particle number, and
harmonic vibrational energy in nonlinear wave and lattice systems.
The associated equilibrium correlations also provide inputs for
descriptions of relaxation and transport.

\section*{Acknowledgments}
This work was supported by the National Natural Science Foundation of China (Grant Nos. 12247106, 12465010, and 12505052). During the preparation of this manuscript, the authors used ChatGPT (OpenAI) to assist in working through and organizing intermediate algebraic steps in several lengthy derivations and in improving the presentation of complex formulas. All mathematical results and scientific conclusions were independently verified by the authors, who take full responsibility for the content of this work.

\bibliographystyle{unsrt}
\bibliography{References}
\end{document}